\documentclass[aps,prb,twocolumn,showpacs, groupedaddress]{revtex4-1}

\usepackage{amsmath,amssymb,bm}
\usepackage{graphicx}
\usepackage[caption=false]{subfig}
\usepackage{bookman}
\usepackage{times}
\usepackage{ctable}
\usepackage{multirow}
\usepackage{floatrow}
\usepackage{wrapfig}
\usepackage{hyperref}

\begin{document}


\title{Ultrathin GaN Nanowires: Electronic, Thermal, and Thermoelectric Properties}
\author{A. H. Davoody}\email[Email: ]{davoody@wisc.edu}
\author{E. B. Ramayya}\altaffiliation[Currently with ]{Intel Corporation, Hillsboro, OR}
\author{L. N. Maurer}
\author{I. Knezevic}\email[Email: ]{knezevic@engr.wisc.edu}
\affiliation{Department of Electrical and Computer Engineering, University of Wisconsin--Madison, Madison, WI 53706-1691, USA}

\begin{abstract}
We present a comprehensive computational study of the electronic, thermal, and thermoelectric (TE) properties of gallium nitride nanowires (NWs) over a wide range of thicknesses (3--9 nm), doping densities ($10^{18}$--$10^{20}$ cm$^{-3}$), and temperatures (300--1000 K). We calculate the low-field electron mobility based on ensemble Monte Carlo transport simulation coupled with a self-consistent solution of the Poisson and Schr\"odinger equations. We use the relaxation-time approximation and a Poisson-Schro\"dinger solver to calculate the electron Seebeck coefficient and thermal conductivity.  Lattice thermal conductivity is calculated using a phonon ensemble Monte Carlo simulation, with a real-space rough surface described by a Gaussian autocorrelation function. Throughout the temperature range, the Seebeck coefficient increases while the lattice thermal conductivity decreases with decreasing wire cross section, both boding well for TE applications of thin GaN NWs. However, at room temperature these benefits are eventually overcome by the detrimental effect of surface roughness scattering on the electron mobility in very thin NWs. The highest room-temperature $ZT$ of 0.2 is achieved for 4-nm-thick NWs, while further downscaling degrades it. In contrast, at 1000 K, the electron mobility varies weakly with the NW thickness owing to the dominance of polar optical phonon scattering and multiple subbands contributing to transport, so $ZT$ increases with increasing confinement, reaching 0.8 for optimally doped 3-nm-thick NWs. The $ZT$ of GaN NWs increases with increasing temperature beyond 1000 K, which further emphasizes their suitability for high-temperature TE applications.
\end{abstract}
\date{\today}
\maketitle

\section {Introduction}
Thermoelectric (TE) devices for clean, environmentally-friendly cooling and power generation are a topic of considerable research activity. \cite{MajumdarS04,Majumdar09} The TE figure of merit, determining the efficiency of a TE device, is defined as $ZT = S^2\sigma T/\kappa$, where $S$ is the Seebeck coefficient (also known as thermopower), $\sigma$ is electrical conductivity, $\kappa$ is thermal conductivity, and $T$ is the operating temperature. Highly doped semiconductors are the materials with the highest $ZT$, because heat in semiconductors is carried mostly by the lattice ($\kappa\approx\kappa_{\mathrm{l}}$), so electronic and thermal transport are largely decoupled; therefore, the power factor, $S^2\sigma$, and thermal conductivity can, in principle, be separately optimized. \cite{MajumdarS04,DiSalvo99,SlackCRC} In order to improve $ZT$, we need to increase the power factor and reduce thermal conductivity. $ZT > 3$ is needed to replace conventional chlorofluorocarbon coolers by TE coolers, but increasing it beyond $1$ has been a challenge.\cite{MajumdarS04}

Nanostructuring has the potential to both enhance the power factor and reduce the thermal conductivity of TE devices.\cite{Shakouri06,VineisAdvMat10} The Seebeck coefficient and the power factor could be higher in nanostructured TE devices than in bulk owing to the density-of-states (DOS) modification, as first suggested by Hicks and Dresselhaus. \cite{HicksPRB93,HicksPRB93QWell} While this effect is expected to be quite pronounced in thin nanowires (NWs), where the DOS is highly peaked around one-dimensional (1D) subband energies, \cite{VoNanoLett08,JeongJAP10,KimJAP09,KimJAP11,NeophytouJEM10,NeophytouPRB11} surface roughness scattering (SRS) of charge carriers counters the beneficial DOS enhancement.\cite{Ramayya12} The field effect has also been shown to enhance the power factor in nanostructures, \cite{LiangNanoLett2009_FieldEffectNanowires,RyuPRL10} as it provides carrier confinement and charge density control without the detrimental effects of carrier-dopant scattering. Moreover, nanostructured obstacles efficiently quench heat conduction, as demonstrated on materials with nanoscale inclusions of various sizes,\cite{VenkatSubNAT01,Harman02,SnyderNatMat08} which scatter phonons of different wavelengths, and on rough semiconductor  nanowires, in which boundary roughness scattering of phonons reduces lattice thermal conductivity by nearly two orders of magnitude.  \cite{BoukaiNAT08,HochbaumNAT08,LimNL2012_surfaceroughnessSiNWs}

Power generation based on TE energy harvesting requires materials that have high thermoelectric efficiency and thermal stability at high temperatures, as well as chemical stability in oxide environments. \cite{Kucukgok2013_MRSProceedings} Bulk III-nitrides fulfill these criteria and have been receiving attention as potential high-temperature TE materials. \cite{Sztein09,Kaiwa07,Pantha08,Yamaguchi03,Tong09,Sztein2013JAP} Bulk GaN, in particular, has excellent electron mobility, but, like other tetrahedrally bonded semiconductors, it also has high thermal conductivity,\cite{Zou02} so its overall TE performance is very modest ($ZT=0.0017$ at 300 K and $0.07$ at 1000 K, as reported by Liu and Balandin \cite{Liu05,Liu005}). Recently, Sztein \emph{et al.} \cite{Sztein2013JAP} have shown that alloying with small amounts of In can considerably enhance the TE performance of bulk GaN. Here, we explore a different scenario: considering that nanostructuring, in particular fabrication of quasi-1D systems such as NWs, has been shown to raise the $ZT$ of other semiconductors, \cite{BoukaiNAT08,HochbaumNAT08,LimNL2012_surfaceroughnessSiNWs} it is worth asking how well GaN NWs could perform in high-temperature TE applications. \cite{Lee09APL_ThermopowerGaN_NWs} There have been a number of advances in the GaN NW growth and fabrication, \cite{Kuykendall2003NL_TriangularNWs,Wang06_GaNGrowthNanotechnology,WangTalin2006_Nanotechnol,Huang02,Simpkins07_DiameterControlGaN_NWs} as well as their electronic characterization, \cite{Huang02,Motayed2007APL,Talin2010SST,CHANG2006_ElecTransport_GaNInN_NWs,Talin09JAP} but very few studies of GaN NWs for TE applications.\cite{Lee09APL_ThermopowerGaN_NWs}

In this paper, we theoretically investigate the suitability of rough $n$-type GaN NWs for high-temperature TE applications. To that end, we simulate their electronic, thermal, and TE properties over a wide range of thicknesses (3--9 nm), doping densities ($10^{18}$--$10^{20}$ cm$^{-3}$), and temperatures (300--1000 K). Electronic transport is simulated using ensemble Monte Carlo (EMC) coupled with a self-consistent Schr\"{o}dinger -- Poisson solver. The electronic Seebeck coefficient and thermal conductivity are calculated by solving the Boltzmann transport equation (BTE) under the relaxation-time approximation (RTA). Lattice thermal conductivity is calculated using a phonon ensemble Monte Carlo simulation, with a real-space rough surface described by a Gaussian autocorrelation function. Throughout the temperature range, the Seebeck coefficient increases while the lattice thermal conductivity decreases with decreasing wire cross section. At room temperature these benefits are eventually overcome by the detrimental effect of SRS on the electron mobility, so the peak $ZT=0.2$ is achieved at 4 nm, with further downscaling lowering the $ZT$. At 1000 K, however, the electron mobility varies very weakly with the NW thickness owing to the dominance of polar optical phonon scattering and multiple subbands contributing to transport, so $ZT$ keeps increasing with increasing confinement, reaching 0.8 for 3-nm-thick NWs at 1000 K and for optimal doing. The $ZT$ of GaN NWs increases with temperature past 1000 K, which highlights  their suitability for high-temperature TE applications.

This paper is organized as follows: the model used to calculate the electron scattering rates is explained in Sec. \ref{sec:electronic}, followed by a discussion of the simulation results for the electron mobility (Sec. \ref{sec:Mobility}) and the Seebeck coefficient (Sec.  \ref{sec:Seebeck}) as a function of wire thickness, doping density, and temperature. In Section  \ref{sec:thermal}, we discuss the phonon scattering models used in this paper and then show the calculated values of phononic and electronic thermal conductivity. In Sec. \ref{sec:zt}, we show the calculation of the thermoelectric figure of merit and discuss its behavior. We conclude with a summary and final remarks in Sec. \ref{sec:conc}.

\section{Electronic Transport}
\label{sec:electronic}

Bulk GaN can crystalize in zincblende or wurtzite structures, the latter being more abundant. In bulk wurtzite GaN, the bottom of the conduction band is located at the $\Gamma$ point. The next lowest valley, located at $M$, is about 1.2 eV higher than $\Gamma$, \cite{BloomGaNBandstructure,Suzuki95} so it does not contribute to low-field electron transport. The electron band structure in the $\Gamma$-valley can be approximated as non-parabolic,  $\mathcal{E}(\textbf{k})\left(1+\alpha \mathcal{E}(\textbf{k})\right)=\hbar^2|\textbf{k}|^2/2m^*$, where $\alpha=0.189$ eV$^{-1}$ is the non-parabolicity factor and $m^*=0.2\,m_0$ is the isotropic electron effective mass, \cite{Albrecht98} given in the units of $m_0$, the free-electron rest mass.

Wurtzite GaN NWs are usually grown \cite{Wang06_GaNGrowthNanotechnology} or etched vertically,  \cite{Frajtag2012_MasklessEtchingGaNWires} along the bulk crystalline \textit{c}-axis, and can have triangular, hexagonal, or quasi-circular cross-sections, depending on the details of processing.\cite{Kuykendall2003NL_TriangularNWs,Wang06_GaNGrowthNanotechnology,WangTalin2006_Nanotechnol,Huang02,Simpkins07_DiameterControlGaN_NWs} In silicon, simulation of electronic transport in rough cylindrical, square, and atomistically realistic nanowires yields results that are remarkably close to one another, both qualitatively and quantitatively, when these differently shaped wires have similar cross-sectional feature size and similar edge roughness features; for instance, the electron mobility in a rough cylindrical wire of diameter equal to 8 nm \cite{Jin07} is very close to the mobility in a square NW with an 8-nm side.\cite{Ramayya08}  Therefore, in order to simplify the numerical simulation of electron and phonon transport in GaN NWs, we consider a square cross section, with the understanding that the wire thickness or width stands in for a generic characteristic cross-sectional feature size.

In the \textit{n}-type doped square GaN NWs considered here, electrons are confined to a square quantum well in the cross-sectional plane and are free to move along the wire axis. In the envelope function approximation, the three-dimensional (3D) electron wave functions have the form  $\psi_{n,k_x}(x,y,z) =\phi_n(y,z)\times\exp(\pm i k_x x)$, where $\phi_n(y,z)$ is a  two-dimensional (2D) wavefunction of the $n$-th 2D subband calculated from the Schr\"odinger-Poisson solver. The corresponding electron energy is $\mathcal{E}_{n,k_x} = \mathcal{E}_n+ \frac{\sqrt{1+4\alpha k_x^2/2m^*}-1}{2\alpha}$, where $\mathcal{E}_n$ is the bottom-of-subband energy. The 2D Poisson's and Schr\"odinger equations are solved in a self-consistent loop: Poisson's equation gives the Hartree approximation for the electrostatic potential, which is used in the Schr\"odinger equation to calculate electronic wave functions and energies in the cross-sectional plane; electronic subbands are then populated to calculate the carrier density and fed back into the Poisson solver. More details regarding the numerical procedure can be found in Refs. \onlinecite{Ramayya08,RamayyaJCEL10}.

Electrons in GaN NWs scatter from acoustic phonons, impurities, surface roughness, polar optical phonons (POP), and the piezoelectric (PZ) field. \cite{Foutz99} GaN NWs have only a few monolayers of native oxide around them; \cite{Prabhakaran96APL,Watkins99APL} therefore, for the purpose of SRS, we will treat them as bare (i.e. surrounded by air). The scattering rates are calculated using Fermi's golden rule. \cite{Lundstrom00} The constants used in the scattering rate calculations are taken from Refs. \onlinecite{Yamakawa09, Foutz99, Lagerstedt79}, and shown in Table \ref{tab:tab1}.

\begin{table}[!]
\begin{ruledtabular}
    \begin{tabular}{ l c c c r}
        Parameter &   & Value & Units & Ref.\\
        \hline
        Deformation potential & $\Xi_{ac}$ & 8.30 & eV &  [\onlinecite{Foutz99}]\\
        Mass density & $\rho$ & 6.15 & $\mathrm{g/cm^3}$ &  [\onlinecite{Foutz99}]\\
        Longitudinal sound velocity & $\upsilon_s$ & 6.56 & $10^3 \mathrm{m/s}$ &  [\onlinecite{Foutz99}]\\
        Lattice constant & $a$ & 3.189 & $\mathrm{\AA}$ & [\onlinecite{Lagerstedt79}]\\
        Lattice constant & $c$ & 5.185 & $\mathrm{\AA}$ & [\onlinecite{Lagerstedt79}]\\
        Optical phonon energy & $E_0$ & 91.2 & meV & [\onlinecite{Foutz99}]\\
        Effective mass &   & 0.2 & - & [\onlinecite{Foutz99}]\\
        Static dielectric constant & $\epsilon_{0}$ & 8.9 & - & [\onlinecite{Foutz99}]\\
        High-frequency diel. constant  & $\epsilon_{\infty}$ & 5.35 & - & [\onlinecite{Foutz99}]\\
        & $e_{15}$ & -0.3 & - & [\onlinecite{Yamakawa09}]\\
        & $e_{31}$ & -0.33 & - & [\onlinecite{Yamakawa09}]\\
        & $e_{33}$ & 0.65 & - & [\onlinecite{Yamakawa09}]\\
        & $c_L$ & $2.65\times10^{11}$ & Pa & [\onlinecite{Yamakawa09}]\\
        & $c_T$ & $4.42\times10^{10}$ & Pa & [\onlinecite{Yamakawa09}]\\
    \end{tabular}
    \caption{GaN material parameters}\label{tab:tab1}
    \end{ruledtabular}
\end{table}

The acoustic phonon scattering rate from subband $n$ to subband $m$ is given by \cite{Ramayya08}
\begin{subequations}\begin{equation}\label{AcousEl}
    \Gamma_{nm}^{ac}(k_x)=\frac{\Xi_{ac}^2 D_{nm}k_BT\sqrt{2m^*}}{2\hbar\rho\upsilon_s^2}\frac{(1+2\alpha\mathcal{E}_f)}{\sqrt{\mathcal{E}_f(1+\alpha\mathcal{E}_f)}}\Theta(\mathcal{E}_f),
\end{equation}
where $\Xi_{ac}$ and $\upsilon_s$ are the deformation potential and the sound velocity, respectively. $\Theta(\mathcal{E}_f)$ is the Heaviside step function and the electron kinetic energy in the final state, $\mathcal{E}_f$, is given by
\begin{equation}\label{Ef}
\mathcal{E}_f=\mathcal{E}_n-\mathcal{E}_m+\frac{\sqrt{1+4\alpha k_x^2/2m^*}-1}{2\alpha}.
\end{equation}
$D_{nm}$ is an overlap integral of the form
\begin{equation}
    D_{nm}=\int|\phi_n(y,z)|^2|\phi_m(y,z)|^2dy\;dz.
\end{equation}
\end{subequations}

\noindent We used a degenerate Thomas-Fermi screening model to calculate the impurity scattering rates. \cite{Kosina98} The impurity scattering rate from subband $n$ to subband $m$ is given by \cite{Ramayya10}
\begin{eqnarray}
    \Gamma_{nm}^{imp}(k_x)&=&\frac{Z^2e^4N_d\sqrt{m^*}}{16\sqrt{2}\pi^2\hbar^2\epsilon_{0}^2}
    \frac{(1+2\alpha\mathcal{E}_f)}{\sqrt{\mathcal{E}_f(1+\alpha\mathcal{E}_f)}}\nonumber\\
    &\times& \int\int \mathrm{d}R \; I_{nm}^2(q_x^\pm,R),\label{ImpEl}\\
    I_{nm}(q_x,R)&=&\int \phi_n(y,z)\mathrm{K}_0(q_x,R)\phi_m(y,z)\;dy\;dz,\label{ImpInt}\nonumber\\
    \mathrm{K}_0(q_x,R)&=&\int_{-\infty}^{+\infty}\frac{e^{iq_x.x}e^{-\frac{\sqrt{(r-R)^2+x^2}}{L_d}}}{\sqrt{(r-R)^2+x^2}}\;dx,\label{K0}\nonumber
\end{eqnarray}
where $\epsilon_{0}$ is the static dielectric permittivity of GaN, $Z=1$ is the number of free electrons contributed by each dopant atom, $N_d$ is the doping density, and $R$ is the position of the impurity atom in the wire. $q_x=\mid k_x-k_x'\mid$ is the magnitude of the difference between the initial ($k_x$) and final ($k_x'$) wave vector of the electron along the wire.

The SRS rate is calculated based on enhanced Ando's model \cite{Ramayya08,Ando82,Jin07}
\begin{subequations}
\begin{eqnarray}\label{SREl}
    \Gamma_{nm}^{sr}(k_x,\pm)&=&\frac{2\sqrt{m^*}e^2}{\hbar^2}\frac{\Delta^2\Lambda}{2+(q_x^\pm)^2\Lambda^2}\mid F_{nm}\mid^2\nonumber\\
    &\times&\frac{(1+2\alpha\mathcal{E}_f)}{\sqrt{\mathcal{E}_f(1+\alpha\mathcal{E}_f)}}\Theta(\mathcal{E}),
\end{eqnarray}
where $\Delta$ and $\Lambda$ are the rms height and correlation length of the surface roughness.  $q_x^\pm=k_x\pm k_x'$ is the difference between the initial ($k_x$) and final ($k_x'$) wavevector, while the plus and minus signs correspond to forward and backward scattering, respectively. $F_{nm}$ is the SRS overlap integral, defined as
\begin{eqnarray}\label{SRSoverlap}
    F_{nm}&=&\int\int dy\;dz\Big[-\frac{\hbar^2}{e\:W\:m_y}\phi_m(y,z)\frac{\partial^2\phi_n(y,z)}{\partial y^2}\nonumber\\
    &+&\phi_n(y,z)E_y(y,z)\left(1-\frac{y}{W}\right)\phi_m(y,z)\\
    &+&\phi_n(y,z) \left(\frac{\mathcal{E}_m-\mathcal{E}_n}{e}\right)\left(1-\frac{y}{W}\right)\frac{\partial\phi_m(y,z)}{\partial y}\Big].\nonumber
\end{eqnarray}
\end{subequations}
The overlap integral in Eq. (\ref{SRSoverlap}) corresponds to scattering from the top surface of the wire ($y=W$, $W$ being the wire thickness and width). The SRS rate from the bottom surface can be calculated by shifting the origin along the $y$-axis. The SRS rate from the side walls can be calculated by exchanging $y$ and $z$ parameters in Eq. (\ref{SRSoverlap}). $E_y(y,z)$ is the $y$-component of the electric field at $(y,z)$.

A detailed derivation of the POP scattering rates is shown in Appendix \ref{sec:POP}. The electron scattering rate by POPs from subband $n$ to subband $m$ is given by
\begin{subequations}
\begin{eqnarray}\label{POPrate}
    \Gamma_{nm}^{POP}(k_x)&=&\frac{e^4\omega_0\left(\epsilon_\infty^{-1}-\epsilon_0^{-1}\right)}{8\pi}N_0\sqrt{\frac{m^*}{2\hbar^2}}I_{1D}(q_x^{\pm},L_y,L_z)\nonumber\\
    &\times&\frac{1+2\alpha\mathcal{E}_f}{\sqrt{\mathcal{E}_f(1+\alpha\mathcal{E}_f)}}\Theta(\mathcal{E}_f).
\end{eqnarray}
As before, $q_x^{\pm}=k_x\pm k'_x$ is the difference between the initial and final electron wavevectors, with plus (minus) corresponding to forward (backward) scattering. Energy conservation determines the final kinetic energy as $\mathcal{E}_f=\mathcal{E}_n-\mathcal{E}_m+\mathcal{E}_i\pm\hbar\omega_0$, where the plus and minus signs correspond to phonon absorption and emission, respectively. $\mathcal{E}_i$ is the initial electron kinetic energy calculated using the non-parabolic band structure. $\epsilon_{\infty}$ and $\epsilon_{0}$ are the high-frequency and low-frequency (static) dielectric permittivities of GaN, respectively (Table \ref{tab:tab1}). $\omega_0$ is the bulk longitudinal optical phonon frequency. $N_0$ is the number of optical phonons with frequency $\omega_0$, given by the Bose-Einestein distribution $$N_0=\frac{1}{e^{\frac{\hbar\omega_0}{k_BT}}-1}.$$ $I_{1D}$ in Eq. (\ref{POPrate}) is the electron-phonon overlap integral defined as
\begin{equation}\label{POPoverlap}
    I_{1D}(q_x,L_y,L_z)=\int\!\int\frac{1}{q^2}\mid I_{nm}(q_y,q_z)\mid^2dq_ydq_z,
\end{equation}
where $I_{nm}(q_y,q_z)$ is defined in Eq. (\ref{eq:Inm}). The integral is taken over the first Brillouin zone.
\end{subequations}

Scattering rate due to the piezoelectric effect is derived in Appendix \ref{sec:PZ} as
\begin{subequations}
\begin{eqnarray}\label{PZrate}
    \Gamma_{nm}^{PZ}(k_x)&=&\frac{K_{av}^2}{4\pi^2\hbar}\frac{e^2k_BT}{\epsilon_{\infty}}\sqrt{\frac{m^*}{2\hbar^2}}I_{1D}(q_x,L_y,L_z)\nonumber\\
    &\times&\frac{1+2\alpha\mathcal{E}_f}{\sqrt{\mathcal{E}_f(1+\alpha\mathcal{E}_f)}}\Theta(\mathcal{E}_f),
\end{eqnarray}
where $I_{1D}$ is the electron-phonon overlap integral in Eq. (\ref{POPoverlap}), and the final kinetic energy is $\mathcal{E}_f=\mathcal{E}_n-\mathcal{E}_m+\mathcal{E}_i$. $\epsilon_{\infty}$ is the high-frequency effective dielectric constant, and $K_{av}$ is the electromechanical coupling coefficient. For the wurtzite lattice, $K_{av}$ is shown to be \cite{Yamakawa09}
\begin{equation}
    K_{av}^2=\frac{\langle e_l^2\rangle}{\epsilon_{\infty} c_L}+\frac{\langle e_t^2\rangle}{\epsilon_{\infty} c_T},
\end{equation}
where
\begin{eqnarray}
    \langle e_l^2 \rangle &=& \frac{1}{7}e_{33}^2+\frac{4}{35}e_{33}(e_{31}+2e_{15})+\frac{8}{105}(e_{31}+2e_{15})^2,\nonumber \\
    \langle e_t^2 \rangle &=& \frac{2}{35}(e_{33}-e_{31}-e_{15})^2+\frac{16}{35}e^2_{15}\\
    &+&\frac{16}{105}e_{15}(e_{33}-e_{31}-e_{15}).\nonumber
\end{eqnarray}
\end{subequations}

\subsection{Electron Mobility}
\label{sec:Mobility}

In GaN nanowires, measured electrical conductivity shows considerable sensitivity to variations in the wire thickness, doping density, and temperature. \cite{Huang02,Kim02,Stern05,Cha06} (For reference, the low-field electron mobility in bulk GaN doped to $10^{19}$ cm$^{-3}$ is of order 200--300 cm$^2$/Vs at room temperature, \cite{ChinJAP94,BarkerJAP05,Yamakawa09} and drops to 100 cm$^2$/Vs at 1000 K.\cite{ChinJAP94}) Here, we perform a comprehensive set of electronic Monte Carlo simulations in order to analyze the dependence of the electron mobility in GaN NWs on the wire thickness, doping density, and temperature. The calculated electron scattering rates are used in a Monte Carlo kernel to simulate electron transport and compute the electron mobility. In these highly doped NWs, the rejection technique is used to account for  the Pauli exclusion principle. \cite{Lugli86}

Electronic Monte Carlo simulations are typically done with 80,000-100,000 particles over timescales longer than several picoseconds, which is enough time to reliably achieve a steady state. Typical ensemble time step is of order 1 fs (much shorter than the typical relaxation times). To insure transport is diffusive, the wire is considered to be very long, so the electronic simulation is actually not done in real space along the wire. Instead, a constant field and an effectively infinite wire are assumed, and the simulation is done in \textit{k}-space. Surface roughness scattering of electrons from the surface with a given rms roughness and correlation length is accounted for through the appropriate SRS matrix element. Across the wire, the Schr\"{o}dinger and Poisson equations are solved self-consistently. A typical mesh across the wire is  67$\times$67 mesh points for 5--9 nm wires, 57$\times$57 for the 4 nm ones. The mesh is nonuniform and is denser near the wire boundary. More details can be found in Ref. [\onlinecite{Ramayya12}].

\begin{figure}[!]
  \subfloat[ ]{\label{fig:fig1a}\includegraphics[width = 3 in]{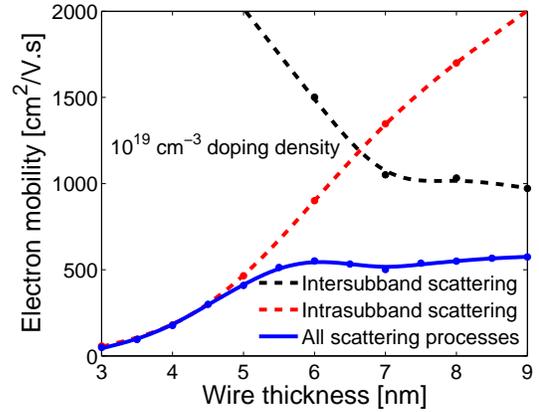}}\\
    \subfloat[ ]{\label{fig:fig1b}\includegraphics[width = 3 in]{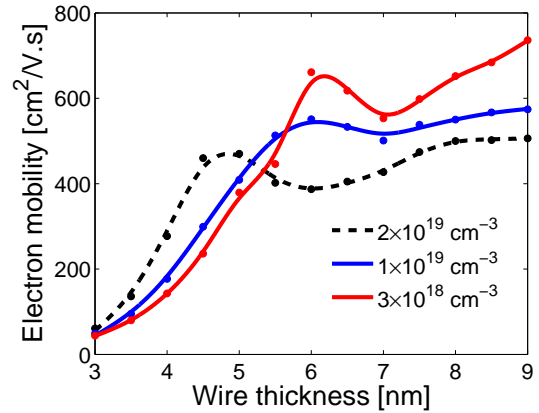}}\\
  \caption{Electron mobility in GaN NWs as a function of thickness. The rms roughness and correlation length of the wire interface are $\Delta=0.3\;\mathrm{nm}$ and $\Lambda=2.5$ nm, respectively. Temperature is $T=300\;\mathrm{K}$. (a) The black dashed curve shows the electron mobility with only intersubband scattering, while the red dashed curve corresponds to intrasubband scattering alone. The blue solid curve shows the net electron mobility, including both inter- and intrasubband scattering processes. (b) Electron mobility in GaN NWs with various doping densities as a function of the NW thickness. }
  \label{fig:fig1}
\end{figure}

First, we discuss the effect of the wire thickness variation on the electron mobility. Figure  \ref{fig:fig1a} shows the electron mobility as a function of the NW thickness for a wire doped to $10^{19}$ cm$^{-3}$. (Doping densities of order $10^{19}$ cm$^{-3}$ are optimal for TE applications in many semiconductors.\cite{MahanBookChapter}) The rms height of the surface roughness is taken to be $\Delta=0.3\;\mathrm{nm}$, as one of the smoothest surfaces reported for GaN crystals.    \cite{Dogan11} The correlation length is assumed to be 2.5 nm, a common value in Si CMOS; we have not been able to find a measured value on GaN systems. The red (black) dashed curve shows the electron mobility when only intrasubband (intersubband) scattering is allowed. The intersubband electron scattering processes are dominant in thicker wires. The intersubband scattering rate decreases with decreasing thickness, as the subband spacing increases. Electrons have higher intrasubband scattering rates in thin wires (red dashed curve), in which the SRS overlap integrals are greater.

Figure \ref{fig:fig1b} shows the electron mobility as a function of the wire thickness for various wire doping densities. Electron mobility has a peak, followed by a dip, around the wire thickness in which the transition from mostly intrasubband to mostly intersubband scattering happens. The dip  in the mobility curve corresponds to the onset of significant intersubband scattering between the lowest two subbands, i.e. the energy difference between the first and second subband bottoms exceeds the polar optical phonon energy. As we can see in Fig. \ref{fig:fig1b}, varying the doping density moves this transition point between mostly inersubband and mostly intrasubband scattering regimes. In thick NWs, similar to bulk, increasing the doping density causes more electron scattering with ionized dopants and the electron mobility decreases. However, for thinner wires the behavior is more complicated and we discuss it in more detail in the next few paragraphs.

\begin{figure}[!]
  \centering
   \includegraphics[width=3 in]{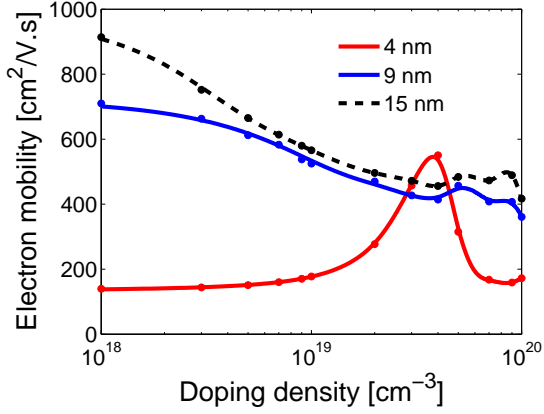}\\
  \caption{Electron mobility in GaN NWs as a function of doping density. The rms roughness and correlation length of the wire interface are $\Delta=0.3\;\mathrm{nm}$ and $\Lambda=2.5$ nm, respectively. Temperature is $T=300\;\mathrm{K}$.}
  \label{fig:fig2}
\end{figure}

Figure \ref{fig:fig2} shows the electron mobility dependence on the doping density for various wire thicknesses. These results are in good agreement with the experimental measurements of Huang \textit{et al.} \cite{Huang02} for a GaN nanowire FET device of 10 nm thickness. In relatively thick NWs, we observe the expected decrease of the electron mobility with increasing doping density. However, for a NW with a relatively small diameter, the electron mobility shows a more complicated non-monotonic behavior with doping density. Similar behavior has been observed by others in the mobility versus effective field dependence of gated silicon nanostructures, \cite{KotlyarAPL04,RamayyaIEEEN07,JinJAP07} where its origin comes from the interplay of surface roughness and nonpolar intervalley phonon scattering in these confined systems. \cite{JinJAP07} In GaN NWs, strong electron confinement is also key, but POP scattering plays the  dominant role instead. The origin of the peak can be readily grasped by relying on the relaxation-time approximation (RTA) expression for the mobility (here $n,k,s$ are the electron quantum numbers -- the subband index, momentum along the NW, and the spin orientation, respectively):

\begin{eqnarray}
\mu_{\mathrm{RTA}}&=&e\frac{\sum_{n,k,s}\upsilon_n(k)\tau_n(k)\left[-\partial f_0/\partial (\hbar k)\right]}{\sum_{n,k,s}f_0(k)}\nonumber\\
&=&\frac{e}{N_D}\sum_{n,k,s}\upsilon_n^2(k)\tau_n(k)\left(- d f_0/d\mathcal E \right)\label{eq:mobilityRTA}\\
&=&\frac{e}{N_D}\int d\mathcal E\sum_{n} g_n(\mathcal E)\upsilon_n^2(\mathcal E)\tau_n(\mathcal E)\left(- d f_0/d\mathcal E \right)\nonumber
\end{eqnarray}

\begin{figure}[!]
\includegraphics[width = 3 in]{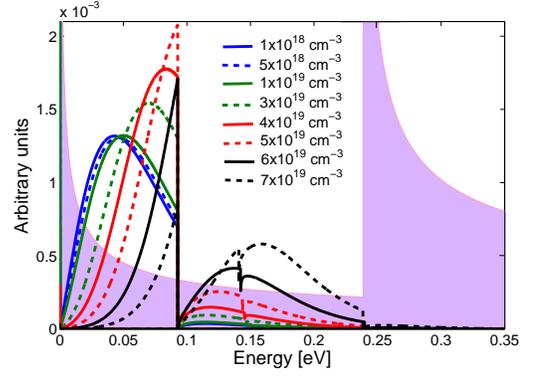}
    \caption{The integrand from the numerator in Eq. (\ref{eq:mobilityRTA}) divided by the doping density $N_D$ for a 4-nm wire and $N_D$ ranging from $10^{18}$ to $7\times 10^{19}$ cm$^{-3}$. The temperature is 300 K. The area under each curve is proportional to the electron mobility. The shaded area corresponds to the density of states, $g(\mathcal E)$. $\mathcal E=0$ is the bottom of the lowest subband.}
  \label{fig:fig3}
\end{figure}

\noindent Here, $f_0(\mathcal E)$ is the Fermi-Dirac distribution function. $g_n(\mathcal E)$ is the density of states, $\tau_n(\mathcal E)$ the lifetime, and $\upsilon_n(\mathcal E)=\left(d \mathcal E/d(\hbar k)\right)_{\mathcal E-\mathcal{ E}_n}$ is the group velocity along the wire for an electron in the $n$-th subband and with the kinetic energy $\mathcal E-\mathcal{E}_n$. We have used the fact that the denominator from the first line of Eq. (\ref{eq:mobilityRTA}), $\sum_{n,k,s}f_0(k)$, equals the electron density, which in turn equals the doping density $N_D$. With increasing doping density, the Fermi level moves up in energy, and the transport window (the energy range where $|df_0/d \mathcal E|$ is appreciable) follows. As $g_n\sim \upsilon_n^{-1}$ in NWs, from the integrand in the numerator of Eq. (\ref{eq:mobilityRTA}) we see that the mobility is determined by the product of $\tau_n(\mathcal E)$ and $\upsilon_n(\mathcal E)$ within the transport window. Figure \ref{fig:fig3} shows the integrand from the numerator of Eq. (\ref{eq:mobilityRTA}) divided by the doping density versus energy, for the 4-nm wire and several doping densities ranging $10^{18}$ cm$^{-3}$ to $7\times 10^{19}$ cm$^{-3}$; the area under each curve is therefore proportional to the mobility for that doping density. (The cumulative density of states, $g(\mathcal E)=\sum_n g_n(\mathcal E)$, is also presented as a lightly shaded area.) Each integrand curve has a steep drop at roughly 91 meV, corresponding to the relaxation time drop due to the onset of intrasubband POP emission for the first subband, and a small dip at about 145 meV, $\sim$91 meV below the second subband bottom, which corresponds to the onset of first-to-second subband intersubband scattering due to POP absorption. For $N_D$ ranging from $10^{18}$ cm$^{-3}$ to $10^{19}$ cm$^{-3}$, the integrand curves overlap, so the areas under them are nearly the same and the mobility is nearly constant. With a further doping density increase, the transport window moves towards the POP emission threshold; mobility reaches its maximal value when the electronic states with high velocities but still below the POP emission threshold are around the middle of the transport window (doping density about $4\times 10^{19}$ cm$^{-3}$). As the density increases futher, the transport window moves into the range of energies with strong intrasubband POP emission and the mobility drops.

\begin{figure}[!]
  \subfloat[ ]{\label{fig:fig4a}\includegraphics[width = 3 in]{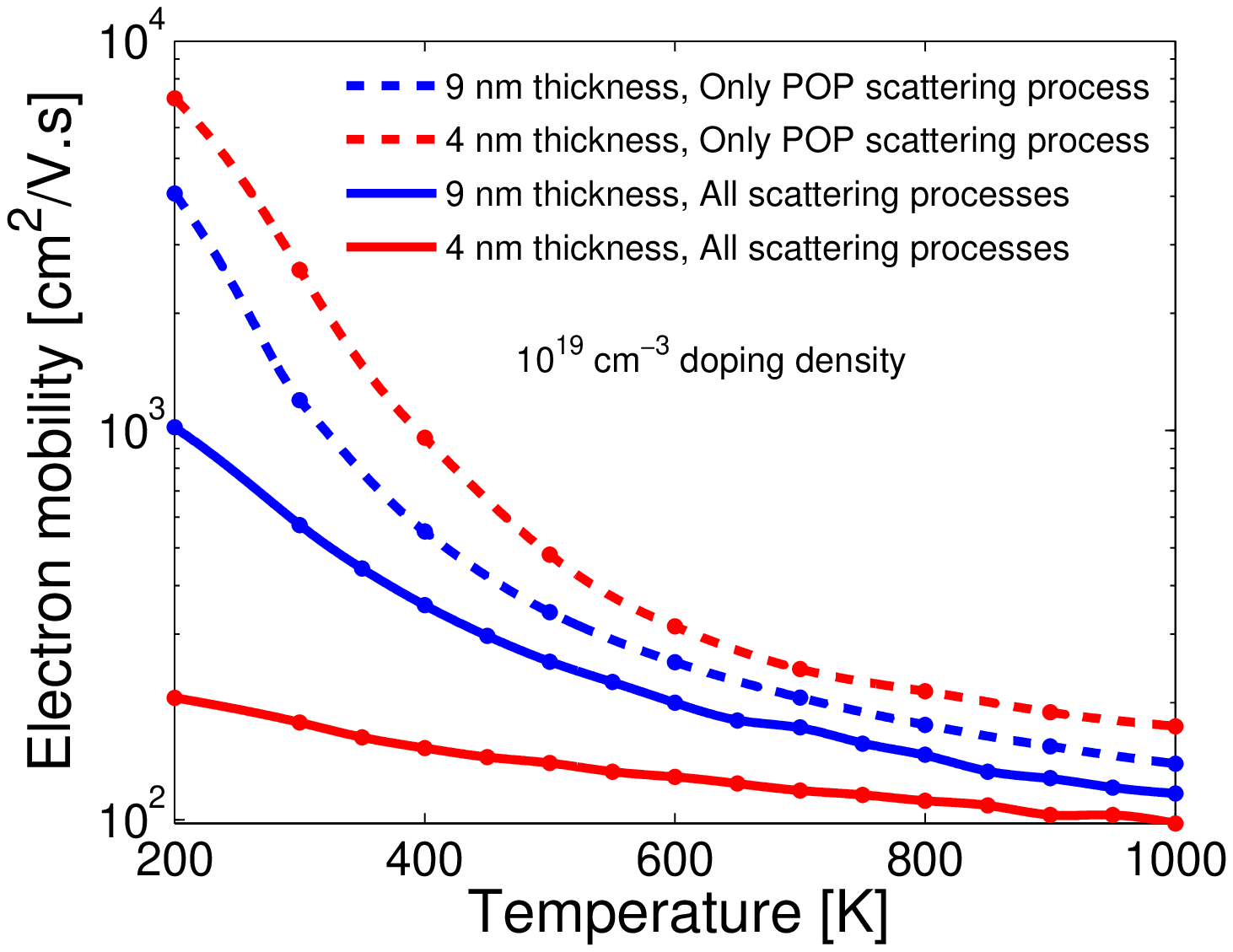}}\\
  \subfloat[ ]{\label{fig:fig4b}\includegraphics[width = 3 in]{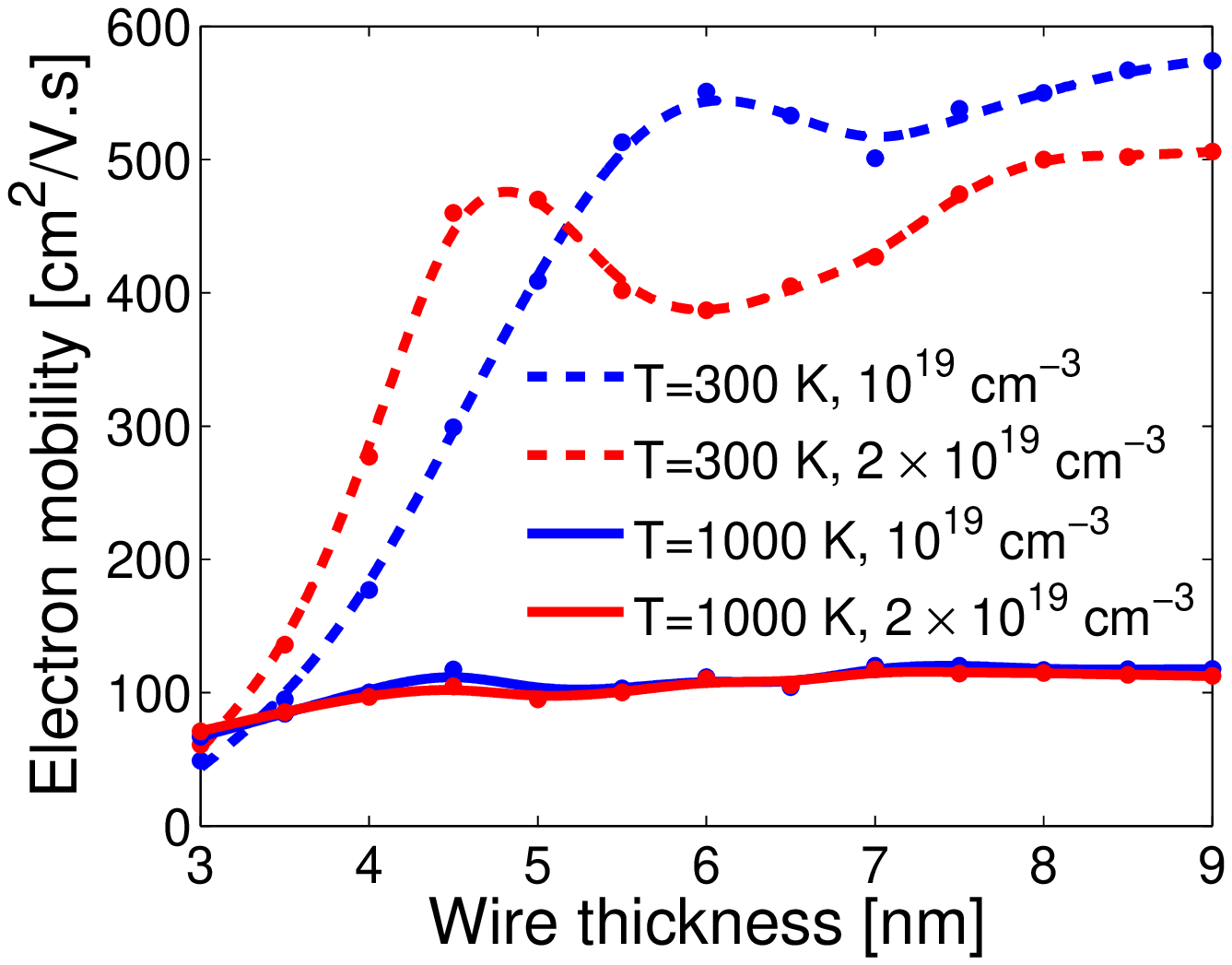}}\\
  \subfloat[ ]{\label{fig:fig4c}\includegraphics[width = 3 in]{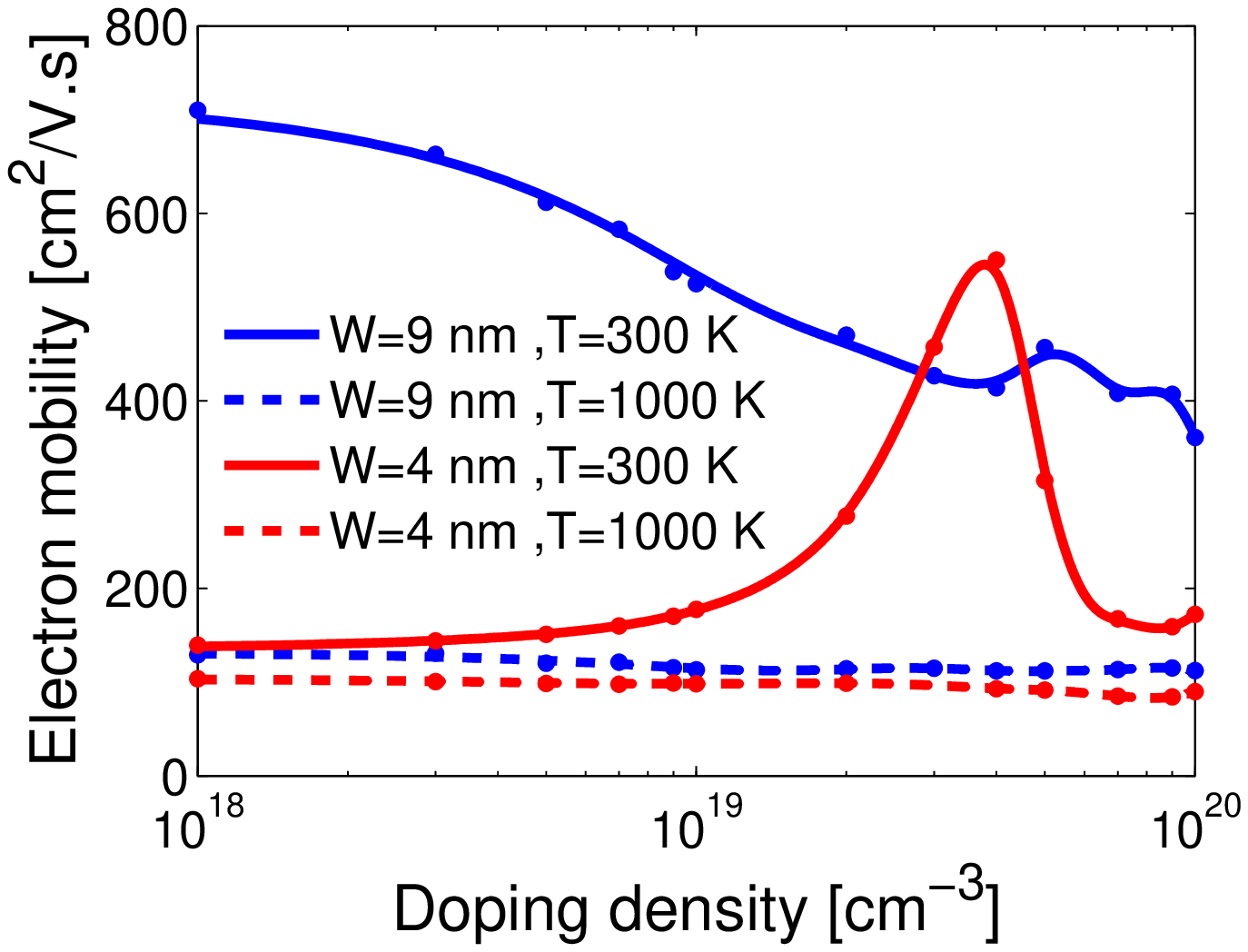}}\\
  \caption{(a) Electron mobility of 4-nm and 9-nm-thick GaN NWs doped to $10^{19}$ cm$^{-3}$ as a function of temperature. (b) Electron mobility as a function of wire thickness at temperatures of $300$ K and $1000$ K and doping densities of $10^{19}$ and $2\times 10^{19}$ cm$^{-3}$. (c) Electron mobility of 4-nm and 9-nm-thick NWs as a function of doping density at $300$ K and $1000$ K. In all three panels, the rms roughness and correlation length of the wire interface are $\Delta=0.3\;\mathrm{nm}$ and $\Lambda=2.5$ nm, respectively.}
  \label{fig:fig4}
\end{figure}

Next, we discuss the effect of a temperature increase on the electron mobility. Figure  \ref{fig:fig4a} shows the electron mobility of 4-nm and 9-nm-thick NWs doped to $10^{19}$ cm$^{-3}$ with only POP scattering and with all scattering mechanisms included. With increasing temperature, the POP scattering rate increase, following the increasing number of polar optical phonons. For temperatures above 600 K, POP scattering becomes the dominant scattering process and governs the rapid decrease of the electron mobility for the thicker, 9-nm NW. In the thinner 4-nm NWs, the mobility is less sensitive to temperature because the greater strength of SRS with respect to POP scattering in thin wires. Figure \ref{fig:fig4b} shows the electron mobility as a function of the wire thickness at 300 and 1000 K for two doping densities, while Fig. \ref{fig:fig4c} presents mobility versus doping density at 300 and 1000 K for 4-nm and 9-nm NWs. At 1000 K,  POP scattering dominates over other mechanisms and the transport window, roughly $3k_BT$ wide, contains a number of subbands; together, these two effects result in flattening of both the mobility vs. wire thickness (Fig. \ref{fig:fig4b}) and the mobility vs. doping density(Fig. \ref{fig:fig4c}) dependencies.

\subsection{The Seebeck Coefficient}
\label{sec:Seebeck}

The Seebeck coefficient (also known as the thermopower) for bulk GaN has a value of 300 -- 400 $\mu$V/K, depending on the sample and the temperature. \cite{Liu05,Kaiwa07,Sztein09} The Seebeck coefficient is a sum of the electronic and the phonon-drag (also known as phononic) contributions.
For GaN NWs, our calculation shows that the phonon-drag Seebeck coefficient is about two orders of magnitude smaller than the electronic one at temperatures of interest, so we henceforth neglect the phonon-drag contribution and equate the total and the electronic Seebeck coefficients. In this section, we discuss the effect of the NW thickness, doping density, and temperature on the Seebeck coefficient.

Based on the 1D BTE using the relaxation-time approximation (RTA), we find the Seebeck coefficient ($S_e$) to be
\begin{equation}
    S_e = -\frac{1}{eT}\frac{\sum_{n}\int{\sqrt{\mathcal{E}}\frac{\partial f_0(\mathcal{E})}{\partial \mathcal{E}}(\mathcal{E}+\mathcal{E}_n-\mathcal{E}_F)\tau_n(\mathcal{E})d\mathcal{E}}}{\sum_{n}\int{\sqrt{\mathcal{E}}\frac{\partial f_0(\mathcal{E})}{\partial \mathcal{E}}\tau_n(\mathcal{E})d\mathcal{E}}},
\end{equation}
where $\mathcal{E}_F$ is the Fermi energy, $f_0(\mathcal{E})$ is the equilibrium Fermi-Dirac distribution, $\tau_n(\mathcal{E})$ is the relaxation time of electron in subband $n$, and $\mathcal{E}_n$ is the energy of the bottom of that subband. Integration over energy is performed from zero to infinity. Note that the Seebeck coefficient is determined by the average excess energy with respect to Fermi energy, $\eta_F=\mathcal{E}+\mathcal{E}_n-\mathcal{E}_F$, carried by electrons in the vicinity of the Fermi level.

\begin{figure}[!]
  \centering
  \subfloat[ ]{\label{fig:fig5a}\includegraphics[width = 3 in]{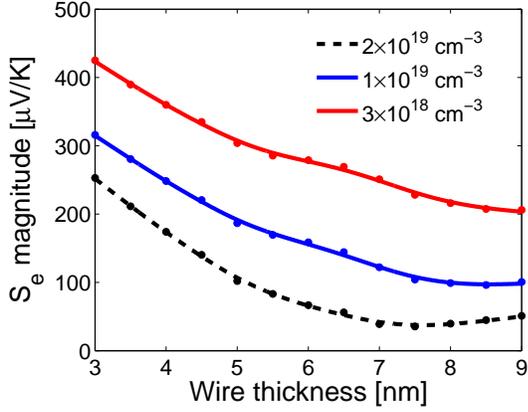}}\\
  \subfloat[ ]{\label{fig:fig5b}\includegraphics[width = 3 in]{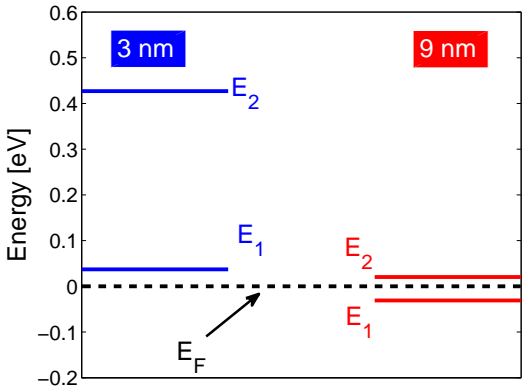}}\\
  \caption{(a) The Seebeck coefficient as a function of NW thickness at T = 300 K and for different doping densities. (b) Positions of the first and second subband bottoms with respect to the Fermi energy for NWs of thickness 3 nm and 9 nm.}
  \label{fig:fig5}
\end{figure}

Figure \ref{fig:fig5a} shows $S_e$ as a function of the wire thickness for various doping densities, which are compatible with the experimental results by Sztein \textit{et al.} \cite{Sztein09}. Decreasing the wire thickness increases the spacing between the subband bottom energies and the Fermi level, which, consequently, increases $\eta_F$ in an average sense (Fig. \ref{fig:fig5b}). The result is a rise in the Seebeck coefficient. For thicker wires, the Fermi level lies between subband bottoms and the interplay between the contributions from different subbands determines the variation of $S_e$. As an example of this interplay, we observe a slight increase in the Seebeck coefficient between the 7-nm and 9-nm-thick NWs at the doping density of $2\times 10^{19}$ cm$^{-3}$.

\begin{figure}[!]
  \centering
  \subfloat[ ]{\label{fig:fig6a}\includegraphics[width = 3 in]{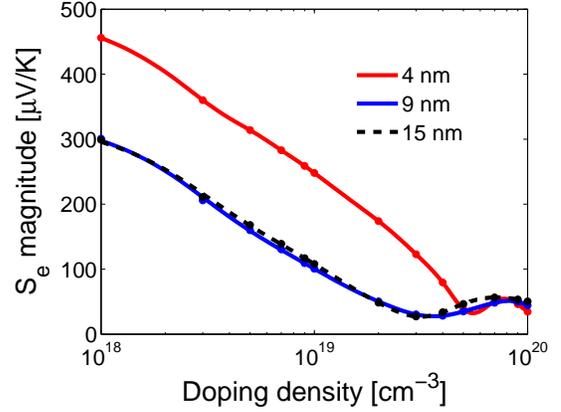}}\\
  \subfloat[ ]{\label{fig:fig6b}\includegraphics[width = 3 in]{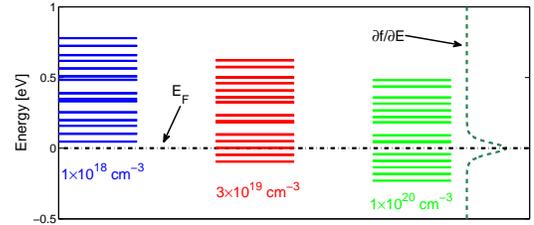}}\\
 \caption{(a) The Seebeck coefficient as a function of doping density for GaN NWs of thickness 4 nm, 9 nm, and 15 nm, at 300 K. (b) Subband energy bottoms with respect to the Fermi energy for a 9-nm-thick wire at three different doping densities. The green dashed line is the negative derivative of the Fermi-Dirac distribution function, $-df_0(\mathcal{E})/d\mathcal{E}$.}
  \label{fig:fig6}
\end{figure}

Figure \ref{fig:fig6a} shows the variation of the Seebeck coefficient with doping density for GaN NWs of different thicknesses. Increasing the doping density means more subband bottoms below the Fermi level (Fig. \ref{fig:fig6b}). This effect results in a high Seebeck coefficient for wires with lower doping densities, for which all subbands are above the Fermi level. In contrast, for degenerately doped wires, the Fermi level typically lies between subbands; $S_e$ is determined by an interplay between the position of the different subbands with respect to the Fermi level (Fig. \ref{fig:fig6b}) and the $S_e$ versus doping density curve is almost flat (Fig. \ref{fig:fig6a}).

\begin{figure}[!]
  \centering
  \subfloat[ ]{\label{fig:fig7a}\includegraphics[width = 3 in]{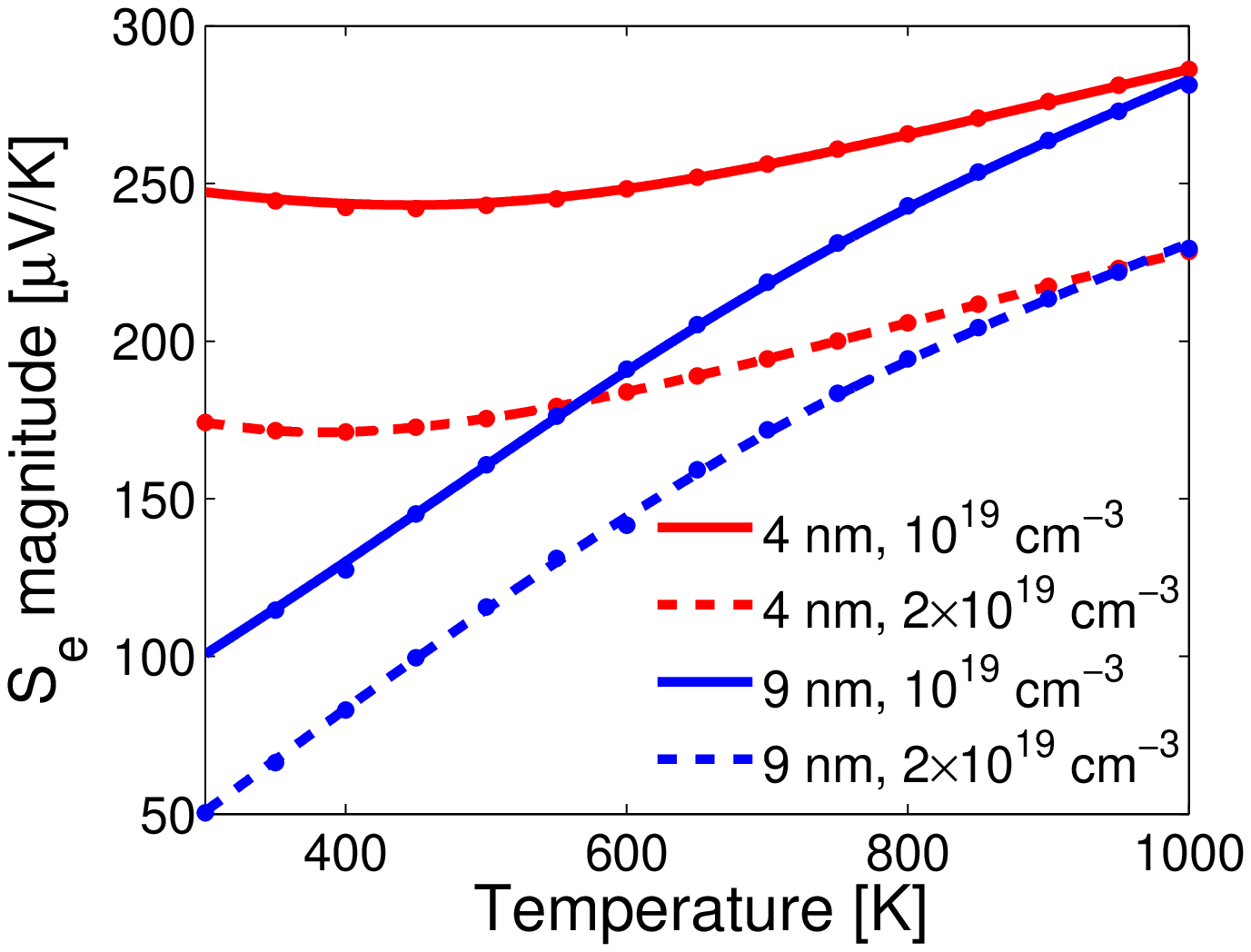}}\\
  \subfloat[ ]{\label{fig:fig7b}\includegraphics[width = 3 in]{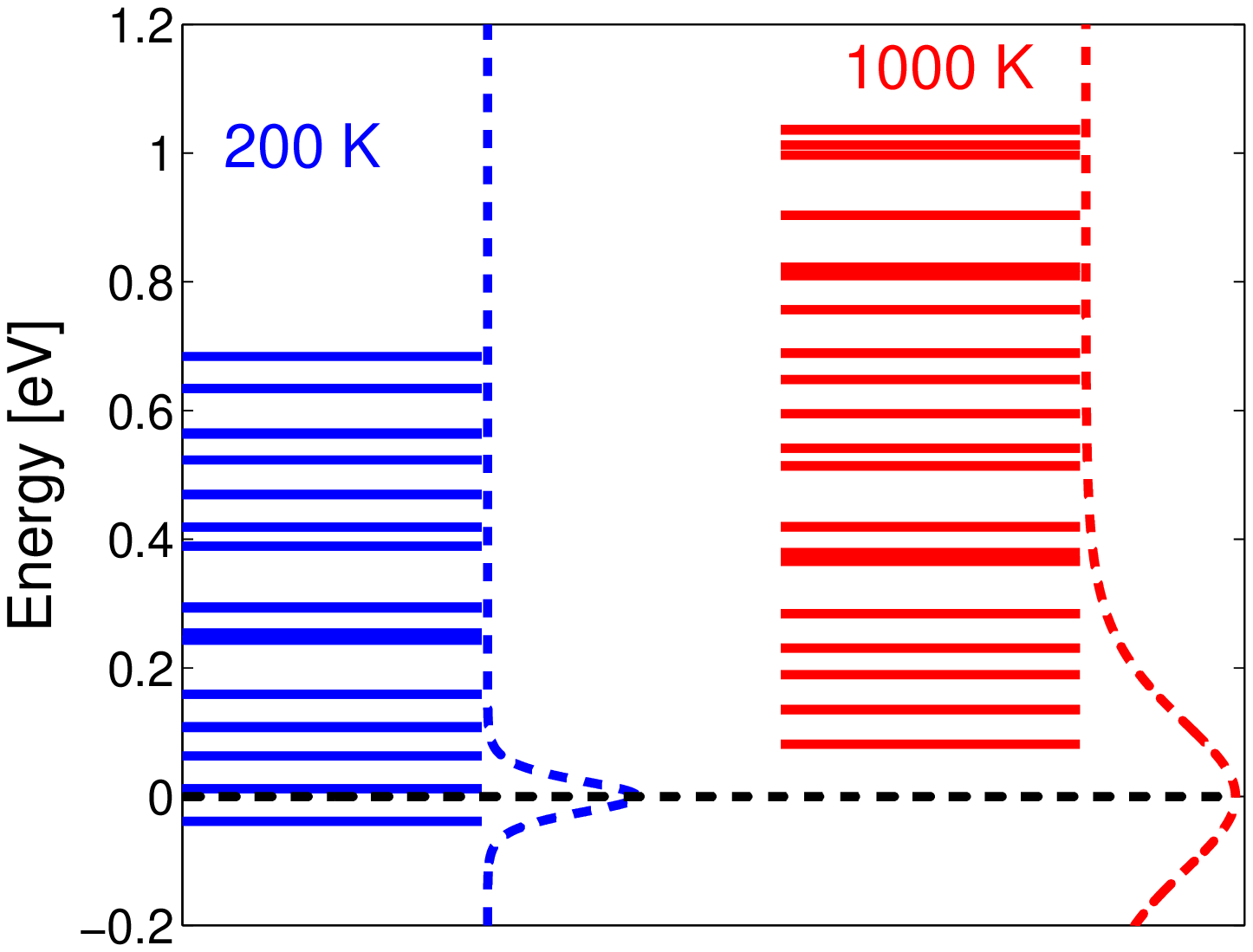}}\\
  \caption{(a) The Seebeck coefficient as a function of temperature for a 9-nm-thick NW with $10^{19}$ cm$^{-3}$ doping density. (b) Subband energy bottoms with respect to the Fermi energy for a 9-nm-thick wire at 200 K and 1000 K. The blue and red dashed lines show the negative of the derivative of the Fermi-Dirac distribution function, $-df_0(\mathcal{E})/d\mathcal{E}$. The black dashed line is the Fermi energy.}
  \label{fig:fig7}
\end{figure}

Figure \ref{fig:fig7a} presents the dependence of the Seebeck coefficient on temperature in 4-nm and 9-nm-thick GaN NWs. A major effect of increasing the temperature is broadening of the Fermi-Dirac distribution function. With increasing temperature, but at a fixed doping density and wire thickness, a given subband will be higher in energy with respect to the Fermi level (thereby contributing more favorably to the Seebeck coefficient) and the energy range for electrons active in electrical conduction  will widen (Fig. \ref{fig:fig7a}). As seen in Fig. \ref{fig:fig7b}, when the temperature is increased from 200 to 1000 K in a 9 nm thick NW doped to $10^{19}$ cm$^{-3}$, the Seebeck coefficient increases by a factor of 3.5.

\begin{figure}[!]
  \centering
  \subfloat[ ]{\label{fig:fig8a}\includegraphics[width = 3 in]{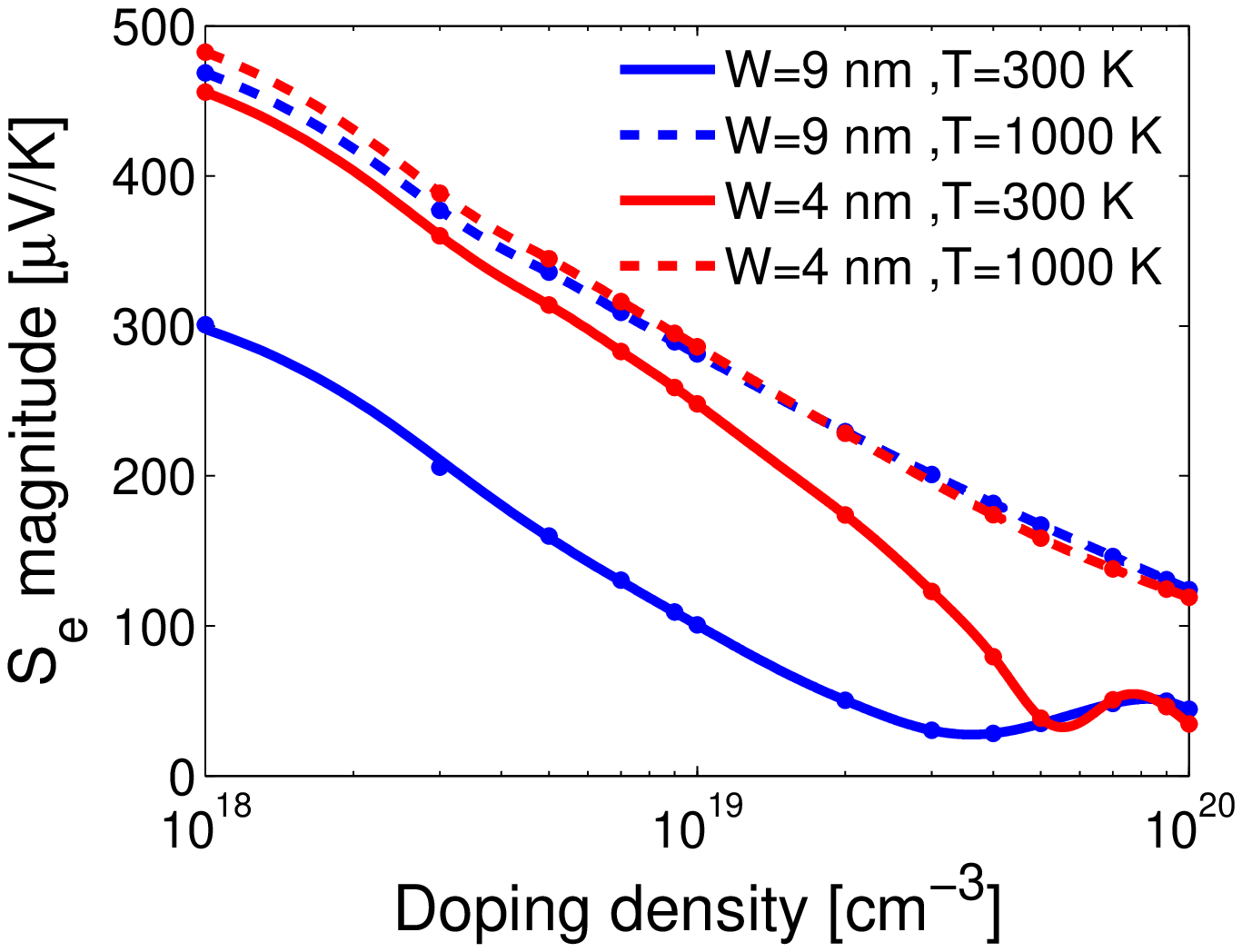}}\\
  \subfloat[ ]{\label{fig:fig8b}\includegraphics[width = 3 in]{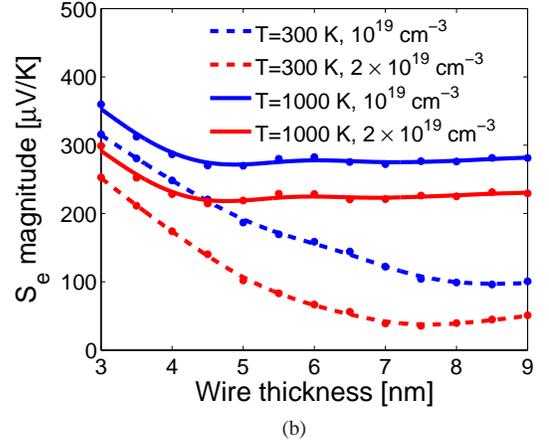}}\\
  \caption{(a) The Seebeck coefficient as a function of doping density for 4-nm and 9-nm-thick NWs at 300 K and 1000 K. (b) The Seebeck coefficient as a function of wire thickness for NWs  doped to $10^{19}$ and $2\times 10^{19}$cm$^{-3}$ at 300 K and 1000 K.}
  \label{fig:fig8}
\end{figure}

Fig. \ref{fig:fig8} shows the Seebeck coefficient as a function of doping density (Fig. \ref{fig:fig8a}) and wire thickness (Fig. \ref{fig:fig8b}) at temperatures 300 K and 1000 K. The Seebeck coefficient increases with increasing temperature, as observed in experiment. \cite{Lee09APL_ThermopowerGaN_NWs}

\section{Thermal Transport}
\label{sec:thermal}

In bulk GaN, most experimental measurements of the thermal conductivity have been done at temperature below 400 K.\cite{Sichel77,Jezowski03} Typical experimental values at room temperature are in the range of 130 -- 200 W/m$\cdot$K; a good survey of the results prior to 2010 was done by AlShaikhi \textit{et al}.\cite{AlShaikhi10} Recent first-principles theoretical calculations by Lindsey \textit{et al}.\cite{LindseyPRL12} give thermal conductivity values for temperatures up to 500 K, with the room-temperature value of about 200 W/m$\cdot$K, in agreement with experiment.\cite{Sichel77,Jezowski03} Based on a theoretical study by Liu and Balandin, \cite{Liu05} the thermal conductivity of bulk GaN at 1000 K  is expected to be about 40 W/m$\cdot$K.

Thermal conductivity in $n$-type NWs comprises two components: phonon (lattice) and electron thermal conductivities. The lattice thermal  conductivity of semiconductor nanowires is expected to be very low, based on theoretical work using  molecular dynamics,\cite{PonomarevaNL07,Papanikolaou08,DonadioNanoLett10,DonadioPRL09}  nonequilibrium Green's functions in the harmonic approximation,\cite{MingoPRB03,WangAPL07,MarkussenPRB09} and the Boltzmann transport equation addressing phonon transport.\cite{ChenJHT05,LacroixPRB05,LacroixAPL06,Mingo04_3,Mingo03}
Here, we calculate the lattice thermal conductivity $\kappa_l$ using the phonon ensemble Monte Carlo (EMC) technique, and electronic thermal conductivity $\kappa_e$ using the RTA. While $\kappa_e$ is much lower than $\kappa_l$ in bulk semiconductors, the two can become comparable in highly doped ultrathin NWs, owing to the reduction of $\kappa_l$ that comes from phonon scattering from rough boundaries and the increase in $\kappa_e$ with increasing doping density. The phonon Monte Carlo method used in this work is explained in detail in Lacroix \emph{et al.}\cite{Lacroix05} and Ramayya \emph{et al.}\cite{Ramayya12} The Monte Carlo kernel simulates transport of thermal energy carried by acoustic phonons; optical phonons are neglected due to their short lifetime and low group velocity. \cite{AlShaikhi10,Kamatagi07} The important acoustic phonon scattering mechanisms are phonon-phonon (normal and Umklapp), mass difference, and surface roughness (boundary) scattering. \cite{AlShaikhi10,Kamatagi07}

We simulate wires of length greater than the typical mean-free path for bulk, in order to properly describe the diffusive transport regime. If the wires are long enough, a linear temperature profile will be obtained along the wire (in contrast with the steplike ballistic transport signature \cite{Ramayya12}). Typical wires in our simulations are 200 nm long. There is no volume mesh, only a surface mesh with typically 1 angstrom mesh cell size, which captures  roughness scattering. Along the wire axis, the wire is divided into cubic segments of the same length as the wire thickness and width. Each segment is assumed to have a well-defined temperature, which is updated during the simulation, as the phonons enter and leave. Energy that is transferred through each boundary between adjacent segments per unit time is recorded and its value averaged along the wire is used to compute the thermal conductivity based on Fourier's law.\cite{Ramayya12}

The normal and Umklapp phonon-phonon scattering rates are calculated using the Holland model. \cite{Holland63} In contrast to the simpler Klemens-Callaway rates, \cite{Callaway59} which assume a single-mode linear-dispersion (Debye) approximation, the Holland rates are more complex as they are specifically constructed to capture the flattening of the dispersive transverse acoustic (TA) modes. \cite{Holland63} For the scattering rate calculation of the TA modes, the zone is split into two regions, such that there is only normal scattering for small wave vectors (roughly up to halfway towards the Brillouin zone edge), while both umpklapp and normal scattering occur for larger wave vectors. Therefore,

\begin{subequations}
\begin{eqnarray}
    \left(\tau_T^N\right)^{-1}&=&B_{TN}\omega T^4,\,  0<\omega<\omega_1,\label{eq:normal-transverse}\\
    \left(\tau_T^U\right)^{-1}&=&\left\{
    \begin{array}{c c}
        0,&\, 0<\omega<\omega_1\\
        \frac{B_{TU}\omega^2}{\sinh(\hbar\omega/k_BT)},& \,\omega_1<\omega<\left(\omega_T\right)_{\mathrm{max}}
    \end{array} \right. \label{eq:umklapp-transverse}
\end{eqnarray}
\end{subequations}
Relaxation rates for longitudinal acoustic phonons are given by
\begin{subequations}
\begin{eqnarray}
    \left(\tau_L^N\right)^{-1}&=&B_{L\:N}\omega^2 T^3,\, 0<\omega<\left(\omega_L\right)_{\mathrm{max}} ,\label{eq:normal-longitudinal}\\
    \left(\tau_L^U\right)^{-1}&=&B_{L\:U}\omega^2 T^3,\, 0<\omega<\left(\omega_L\right)_{\mathrm{max}} .\label{eq:normal-longitudinal}
\end{eqnarray}
\end{subequations}

\noindent Here, $B_{TN}$, $B_{TU}$, $B_{LN}$, and $B_{LU}$ are the constants shown in Table \ref{tab:tab2}, which are calculated by fitting our simulation results for bulk GaN to experimental results of Sichel \emph{et al.} \cite{Sichel77} $\omega_1$ corresponds to the frequency of the transverse branch at $q_{\mathrm{max}}/2$ point, where $q_{\mathrm{max}}$ is the Brillouin zone boundary. \cite{Holland63}

\begin{table}[!]
\begin{ruledtabular}
  \centering
  \begin{tabular}{l  c r}
    Parameter & Value & Units \\
    \hline
    $B_{LN}$ & $1.2\times10^{-24}$ & $\mathrm{s.K}^{-3}$ \\
    $B_{LU}$ & $1.2\times10^{-24}$ & $\mathrm{s.K}^{-3}$ \\
    $B_{TN}$ & $3.2\times10^{-12}$ & $\mathrm{K}^{-4}$ \\
    $B_{TU}$ & $2.08\times10^{-17}$ & s \\
  \end{tabular}
  \caption{Phonon-phonon scattering fitting parameters \cite{Kamatagi07}}
  \label{tab:tab2}
  \end{ruledtabular}
\end{table}

The relaxation rate for mass difference scattering is given by the following expression \cite{Klemens58}
\begin{subequations}
\begin{equation}
    \left(\tau_{\:I}\right)^{-1}=A_i \omega^4 ,\label{eq:impurity}
\end{equation}
where $A_i$ is a sample-dependent constant, given by
\begin{equation}
\label{eq:Ai}
    A_i=\frac{V_0\Gamma}{4\pi\upsilon_s^3}.
\end{equation}
\end{subequations}
Here, $V_0$ is the volume per atom, equal to $V_0=\frac{\sqrt{3}}{8}a_0^2c_0$ for the wurtzite crystal. $\upsilon_s$ is the average phase velocity given by $\upsilon_s^{-1}=\frac{1}{3}[2\upsilon_T^{-1}+\upsilon_L^{-1}]$ under the isotropic phonon dispersion approximation. \cite{Liu005} $\Gamma$ is the constant which indicates the strength of mass difference scattering. It is defined as $\Gamma=\sum_{\substack{i}} f_i[1-(M_i/M)]^2$, where $f_i$ is the fractional concentration of the atoms type $i$ with different mass, $M_i$, in the lattice. $M$ is the average atomic mass, $M = \sum_{\substack{i}} f_iM_i$. The $\Gamma$ parameter due to isotopes for a typical sample is given in Table \ref{tab:tab3}. The $\Gamma$ parameter due to doping with Si to $10^{19}$ cm$^{-3}$ is about $7.5\times 10^{-5}$, an order of magnitude smaller than for isotopes. Mass difference scattering due to dopants becomes comparable to isotope scattering at about $10^{20}$ cm$^{-3}$ doping density; therefore, thermal conductivity has a weak dependence on the doping density. The total relaxation rate is given by $\tau_{\mathrm{tot}}^{-1}=\tau_N^{-1}+\tau_U^{-1}+\tau_I^{-1} $.

\begin{figure}[!]
\centering
\includegraphics[width=5 cm,trim=0cm 0cm 0cm 0cm,clip=true]{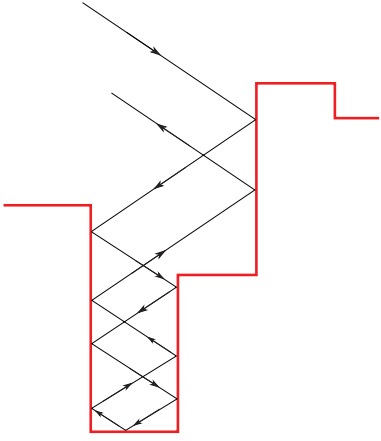}
\caption{A phonon hitting a rough surface spends some time bouncing around, effectively appearing to be localized at the surface.}
\label{fig:fig9}
\end{figure}

\begin{table*}[!]
\begin{ruledtabular}
\begin{center}
\begin{tabular}{l c r}
Parameter       & &Value \\
\hline
Lattice constant &$a_0$ & $3.19$ \AA \\
Lattice constant &$c_0$ & $5.19$ \AA \\
L branch phonon group velocity at point $\Gamma$ & $\upsilon_{L}$ & $7.96\times10^5\;\mbox{cm}/\mathrm{s}$ \\
T branch phonon group velocity at point $\Gamma$ & $\upsilon_{T}$ & $4.13\times10^5\;\mathrm{cm}/\mathrm{s}$ \\
Longitudinal phonon frequency at point $M$ & $f_{L}$ & $9$ THz \\
Transverse phonon frequency at point $M$ & $f_{T}$ & $6.3$ THz \\
Transverse phonon dispersion curve fitting parameter & $c_T$ & $-5.73\times 10^{-8}\; \mathrm{m}^2/\mathrm{s}$ \\
Longitudinal phonon dispersion curve fitting parameter & $c_L$ & $-2.63\times 10^{-7}\; \mathrm{m}^2/\mathrm{s}$ \\
Isotope scattering parameter & $\Gamma$ & $2 \times 10^{-4}$\\
\end{tabular}
\end{center}
\caption{GaN material parameters, from Ref. \onlinecite{Siegle97}. $\Gamma$ is assumed to be for sample 2 in Ref. \onlinecite{Jezowski003}.}
\label{tab:tab3}
\end{ruledtabular}
\end{table*}
Surface roughness scattering is often modeled using the RTA and a specularity parameter that accounts for diffuse scattering at the surface. \cite{Mazumder01,AksamijaPRB10} Here, we have accounted for SRS more realistically by generating a rough surface with specific rms roughness and correlation length. \cite{Ramayya12} When a phonon hits the rough surface, it will reflect specularly at the point of impact; this approach is reminiscent of ray-tracing (see, for instance, Refs. \onlinecite{MooreAPL08,TermentzidisPRB13}). The phonon can undergo multiple reflections before it returns inside the wire (Fig. \ref{fig:fig9}).

We used a quadratic dispersion relationship, $\omega_0=\upsilon_s q_0+c_s q_0^2$, fitted to the experimental data of Ref. \onlinecite{Siegle97}, for transverse and longitudinal bulk phonons. The quadratic dispersion is quite accurate in wurtzite GaN, as shown, for example, by Ma \emph{et al}.\cite{MaJAP13} $\upsilon_s$ is the the sound velocity (i.e. the phonon group velocity at the $\Gamma$ point). The material parameters are listed in Table \ref{tab:tab3}.

\begin{figure}[!]
  \centering
  \subfloat[ ]{\label{fig:fig10a}\includegraphics[width = 3 in]{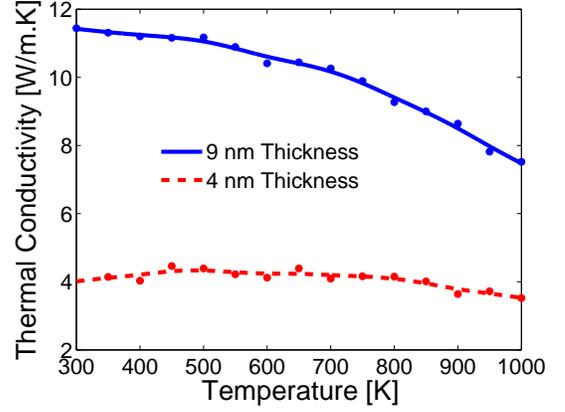}}\\
  \subfloat[ ]{\label{fig:fig10b}\includegraphics[width = 3 in]{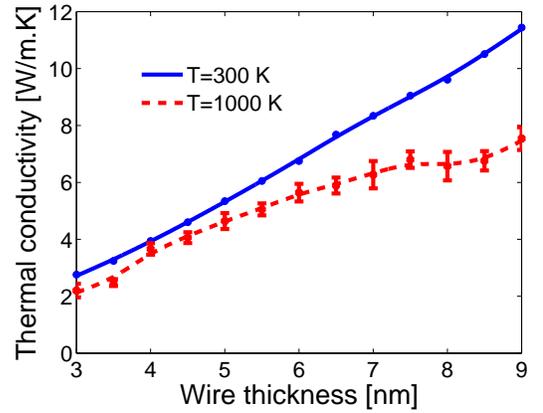}}\\
  \caption{(a) Thermal conductivity of 4-nm and 9-nm-thick GaN NWs as a function of temperature.  (b) Thermal conductivity as a function of NW thickness at 300 K and 1000 K. Roughness rms height is 0.3 nm and the correlation length is 2.5 nm.}
  \label{fig:fig10}
\end{figure}

Figure \ref{fig:fig10} shows thermal conductivity as a function of temperature and wire thickness, for rms roughness of 0.3 nm and a correlation length of 2.5 nm. For these roughness parameters, thermal conductivity in GaN NWs shows a reduction by a factor of 20 with respect to bulk at 300 K, which emphasizes the dominance of SRS in phonon transport over other processes.

The slight waviness in the 1000 K in Fig. 10b is of numerical origin; with increasing temperature the number of real phonons represented by one numerical phonon increases rapidly, which affects accuracy. While the error bars on the 300 K data are too small to be visible, the 1000 K values are of order a few percent (Figure \ref{fig:fig10}b).

\section{Figure-of-Merit Calculation}
\label{sec:zt}

\begin{figure}[!]
  \centering
  \subfloat[ ]{\label{fig:fig11a}\includegraphics[width = 3 in]{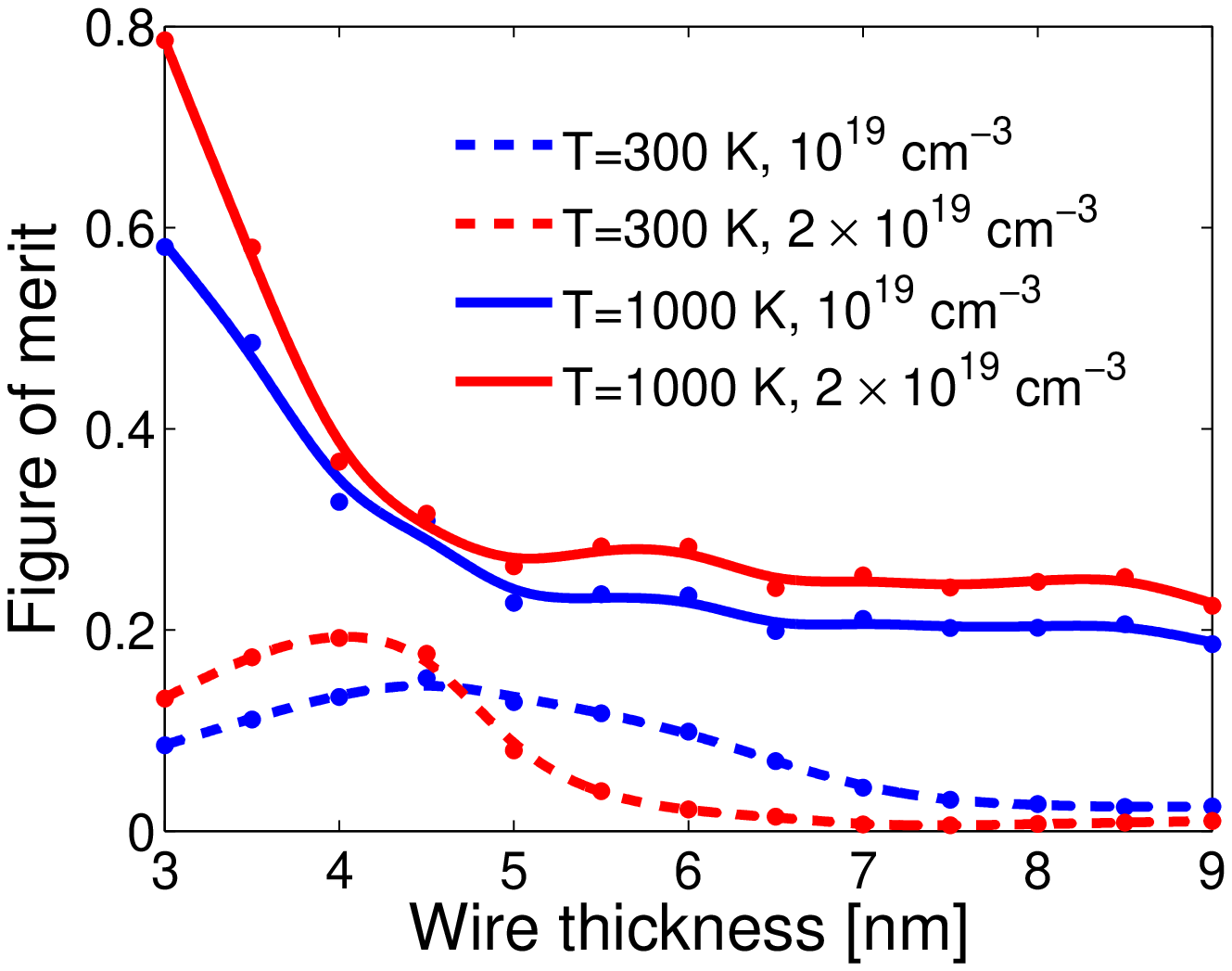}}\\
  \subfloat[ ]{\label{fig:fig11b}\includegraphics[width = 3 in]{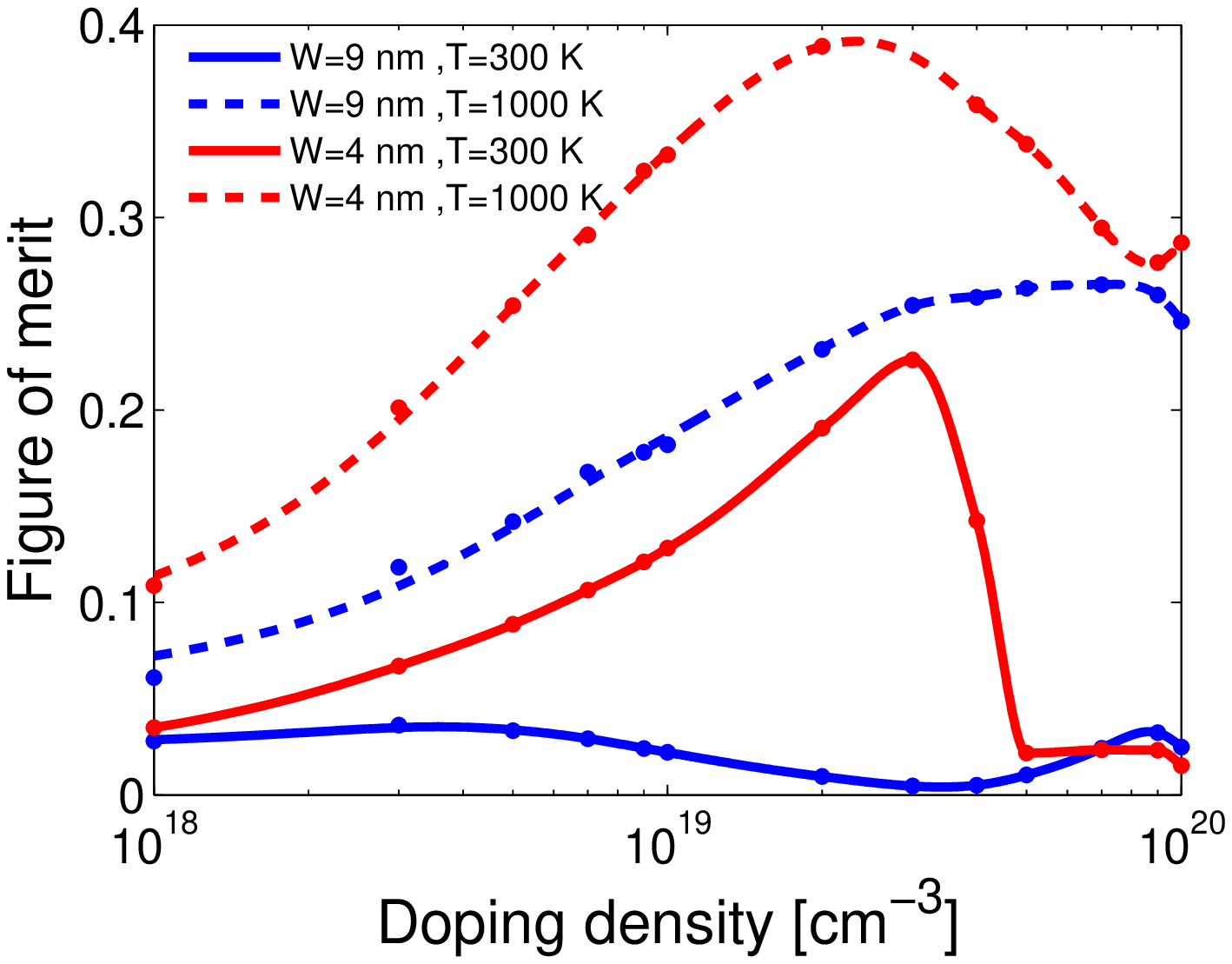}}\\
  \subfloat[ ]{\label{fig:fig11c}\includegraphics[width = 3 in]{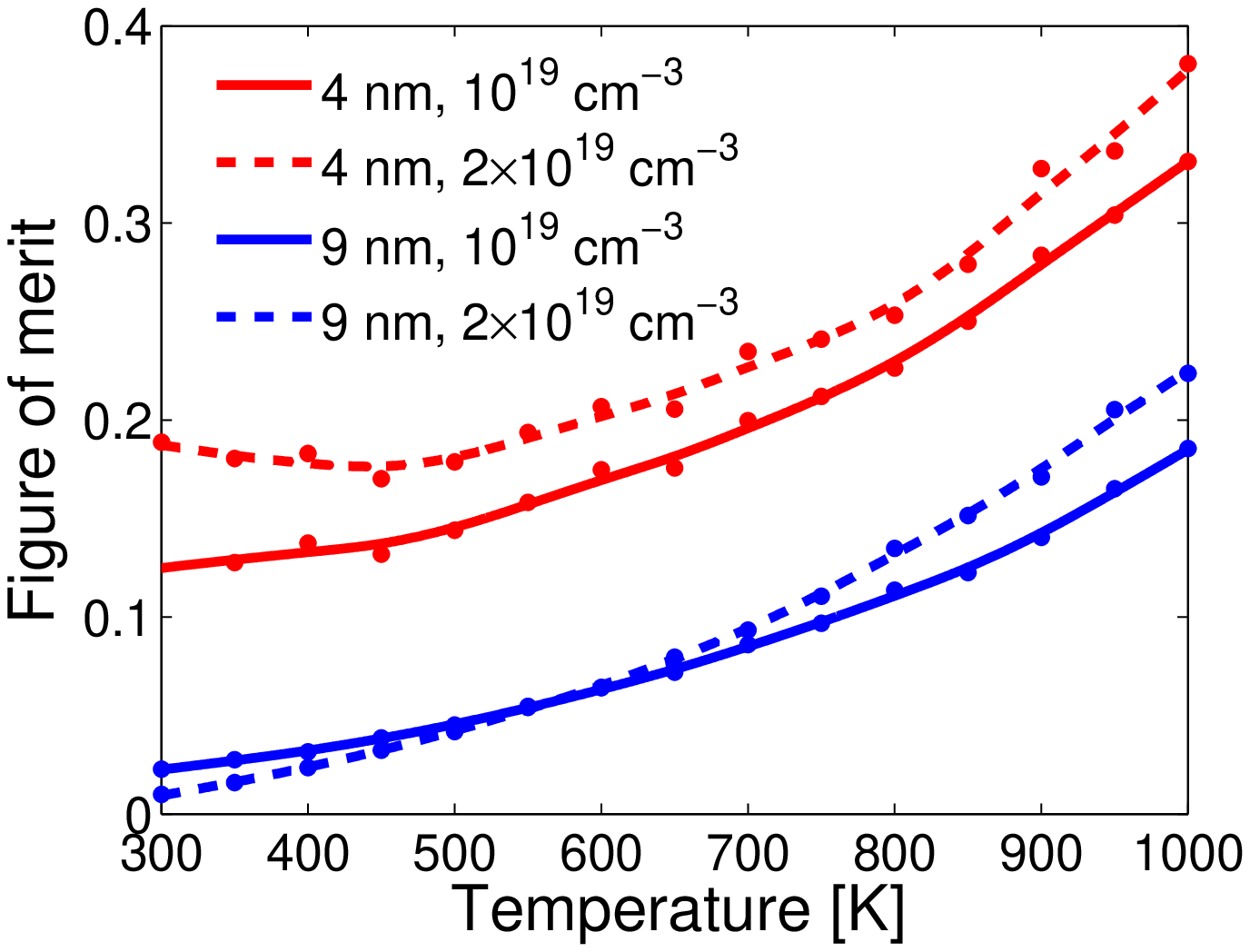}}\\
  \caption{The TE figure of merit of GaN NWs as function of (a) wire thickness (doping density $10^{19}$ and $2\times 10^{19}$ cm$^{-3}$, temperature 300 and 1000 K), (b) doping density (4 and 9 nm thickness, temperatures 300 and 1000 K), and (c) temperature (4 and 9 nm thickness, doping density $10^{19}$ and $2\times 10^{19}$ cm$^{-3}$).}
  \label{fig:fig11}
\end{figure}

Using the calculated electron mobility, Seebeck coefficient, and lattice thermal conductivity, we compute the TE figure of merit. Figure \ref{fig:fig11} shows the variation of room-temperature $ZT$ as a function of wire thickness (Fig. \ref{fig:fig11a}), doping density (Fig. \ref{fig:fig11b}),  and temperature (Fig. \ref{fig:fig11c}).

The highest room-temperature $ZT$ values in GaN NWs of approximately 0.2 are two-orders-of-magnitude greater than the  bulk $ZT$ value of 0.0017 reported by  Liu and Balandin, \cite{LiuBalandinAPL04,Liu005} an increase that stems both from the thermal conductivity reduction  (Fig. \ref{fig:fig10b}) and the Seebeck coefficient increase (Fig. \ref{fig:fig8b}) with decreasing wire thickness. Wires with characteristic cross-sectional features of about 4 nm have the highest $ZT$ values at room temperature; the decrease in $ZT$ with further reduction in thickness comes from the overall detrimental effect of SRS on the electron mobility, which overshadows the beneficial effects of thermal conductivity reduction and $S_e$ increase.

In contrast, at 1000 K, the transport window contains multiple subbands and POP scattering is the dominant scattering mechanism, so the electron mobility is nearly independent of both thickness and doping density. As a result, the $ZT$ of GaN NWs continues to increase with decreasing thickness, and reaches 0.8 in 3-nm-thick  GaN NWs for the $2\times 10^{19}$ cm$^{-3}$ doping density. (For wires thinner than 3 nm, changes in the phonon dispersion and electronic band structure become considerable and atomistic approaches ought to be employed,\cite{Neophytou2011PRB} which may quantitatively change the TE figure of merit.) The $ZT$ of GaN NWs continues to rise with increasing temperature beyond 1000 K, which should ensure efficient energy harvesting with these devices up to high temperatures.

\section{Conclusion}
\label{sec:conc}

We presented a comprehensive computational study of the electronic, thermal, and thermoelectric properties of GaN NWs over a broad range of thicknesses, doping densities, and temperatures.

At room temperature, SRS of electrons in thin GaN NWs competes with polar optical phonon scattering, and results in a decrease of the electron mobility with decreasing thickness. Roughness also decreases thermal conductivity in thin wires, which is beneficial in thermoelectric applications. Reduced wire thickness improves the Seebeck coefficient, which is considerably higher in thin wires over in bulk, owing to the combined effects of the 1D subband density-of-states and the increasing subband separation that follows a reduction in the wire cross section. Cumulatively, reducing the wire cross-sectional features down to 4 nm results in the room-temperature $ZT$ increasing, with a maximum of 0.2 obtained for wires of 4-nm thickness doped to $2\times 10^{19}$ cm$^{-3}$, a two-orders-of-magnitude increase over bulk. Below 4 nm, the room-temperature $ZT$ does not improve with further confinement, as the detrimental surface-roughness-scattering of electrons and the drop in mobility win over the beneficial effects that confinement has on the Seebeck coefficient and thermal conductivity.

At high temperatures, the highest in this study being 1000 K, the electron mobility flattens as a function of thickness, as many subbands start to contribute to transport and POP scattering wins over the temperature-insensitive SRS. The Seebeck coefficient is higher at 1000 K than at 300 K and increases with decreasing wire thickness, although less dramatically than at lower temperatures, while thermal conductivity beneficially decreases with increased confinement. Overall, at 1000 K the thermoelectric figure of merit increases with increasing confinement (i.e. decreasing NW thickness), reaching a value of 0.8 for 3 nm wires.

The $ZT$ of GaN NWs continues to increase with the temperature increasing beyond 1000 K, owing to the negligible minority carrier generation across the large gap, which underscores the suitability of these structures for high-temperature energy-harvesting applications. Extrapolation of the trend would yield  $ZT=1$ at 2000 K for 4-nm-thick NWs.  Further improvements in $ZT$ might be achieved by additional alloy scattering of phonons by introducing In, as demonstrated in Ref. \onlinecite{Sztein2013JAP}. Combined with nanostructuring, InGaN NWs might prove to be a particularly interesting choice for high-temperature power generation.

\section{Acknowledgement}

The authors thank Z. Aksamija and T. Kuech for helpful discussions. This work was primarily supported by the NSF grant No. 1121288 (the University of Wisconsin MRSEC on Structured Interfaces, IRG2, funded A.H.D.), with partial support by the AFOSR grant No. FA9550-09-1-0230 (funded
E.B.R. and I.K.) and by the NSF grant No. 1201311 (funded L.N.M.).

\appendix
\section{Polar Optical Phonon Scattering}
\label{sec:POP}

In wurtzite crystals, there is no clear distinction between the longitudinal and transverse optical phonon modes. Based on careful calculations, Yamakawa \textit{et al.} \cite{Yamakawa09} have shown that electrons have two orders of magnitude higher scattering rates with the LO-like modes than the TO-like ones, so it is sufficient to consider the LO-like modes alone in electronic transport calculations. Furthermore, there is a profound anisotropy in the bulk electron-phonon scattering rate with respect to the electron momentum (see Yamakawa \textit{et al.}, Ref. \onlinecite{Yamakawa09}). As our wires are assumed to be along the wurtzite \textit{c}-axis, we consider only LO-like  phonons interacting with electrons whose initial and final momenta are along the \textit{c}-axis. In this case, there is a single relevant phonon energy, whose value of 91.2 meV is taken after Ref. \onlinecite{Foutz99} and is also given in Table \ref{tab:tab1}.

Here, we show the detailed calculation of the electron-longitudinal polar optical phonon  scattering rate. The electric field due to the propagation of a longitudinal optical phonon is given by
\begin{equation}\label{eq:A1}
    \vec{E}(q)=\sqrt{\frac{\hbar}{2\gamma V \omega_0}}\left(a_qe^{i\vec{q}\cdot\vec{r}}+a_q^\dagger e^{-i\vec{q}\cdot\vec{r}}\right)\cdot\vec{e}_q\, ,
\end{equation}
where $\vec{e}_q$ is the polarization vector, $a_q$ ($a_q^\dagger$) is the phonon creation (annihilation) operator, and $\omega_0$ is the optical phonon frequency. $\gamma$ is the effective interaction parameter given by
\begin{equation}
    \frac{1}{\gamma}=\omega_0^2\left(\frac{1}{\epsilon_{\infty}}-\frac{1}{\epsilon_{0}}\right).
\end{equation}
Here, $\epsilon_{\infty}$ and $\epsilon_{0}$ are the high-frequency and low-frequency dielectric permittivities, respectively.
From Eq. (\ref{eq:A1}), the perturbing Hamiltonian is equal to
\begin{equation}
    \textbf{H}_{pop}=\sum_{\vec{q}} \frac{C}{q}\left(a_qe^{i\vec{q}\cdot\vec{r}}-a_q^\dagger e^{-i\vec{q}\cdot\vec{r}}\right),
\end{equation}
where $C=i\sqrt{\frac{\hbar e^2 \omega_0}{2 V}\left(\frac{1}{\epsilon_{\infty}}-\frac{1}{\epsilon_{0}}\right)}$.

The matrix element for scattering from the initial electronic state $|k_x,n\rangle$ to the final state $|k_x',m\rangle$ is given by
\begin{eqnarray}\label{eq:Mnm}
    M_{nm}(k_x,k_x',q)&=&\langle k_x',m| \textbf{H}_{pop}(q) |k_x,n\rangle \nonumber\\
    &=&\frac{C}{q} \sqrt{N_0+\frac{1}{2}\pm\frac{1}{2}} \\
    &\times&\int \psi_n(y,z)e^{i(q_yy+q_zz)}\psi_m(y,z)\;dy\;dz\nonumber\\
    &\times&     \frac{1}{L_x}\int e^{i(k_x-k_x'\mp q_x)x}dx,\nonumber
\end{eqnarray}
where plus and minus correspong to emission and absorption of POP, respectively. $N_0$ is the number of optical phonons given by the Bose-Einstein distribution function
\begin{equation}
    N_0=\frac{1}{e^{\frac{\hbar \omega_0}{k_B T}}-1}.
\end{equation}

We define the function $I_{nm}(q_y,q_z)$ as
\begin{equation}\label{eq:Inm}
    I_{nm}(q_y,q_z)=\int\left[\psi_n(y,z)e^{i(q_yy+q_zz)}\psi_m(y,z)\right]dy\;dz,
\end{equation}
and the Eq. (\ref{eq:Mnm}) yields
\begin{eqnarray}
    \left|M_{nm}(k_x,k_x',q)\right|^2&=&\frac{|C|^2}{q^2}\left(N_0+\frac{1}{2}\pm\frac{1}{2}\right)\\
    &\times&\left|I_{nm}(q_y,q_z)\right|^2\delta(k_x-k_x'\mp q_x).\nonumber
\end{eqnarray}
According to Fermi's golden rule, the polar optical phonon scattering rate is given by
\begin{equation}\label{eq:gammapop}
    \Gamma_{nm}^{pop}=\frac{2\pi}{\hbar}\sum_{q_{\|},k_x'}\left|M_{nm}(k_x,k_x')\right|^2\delta(\mathcal{E}'-\mathcal{E}\pm\hbar\omega_0).
\end{equation}
By changing the sum to integral we get
\begin{eqnarray}
    \Gamma_{nm}^{pop}(k_x)&=&\frac{|C|^2V}{4\pi^2\hbar}\left(N_0+\frac{1}{2}\pm\frac{1}{2}\right)\\
    &\times& \int dk_x'\int \left|I_{nm}(q_y,q_z)\right|^2dq_y\;dq_z \nonumber\\
    &\times&\delta(k_x-k_x'\mp q_x)\delta(\mathcal{E}'-\mathcal{E}\pm\hbar\omega_0).\nonumber
\end{eqnarray}

Next, we define the overlap integral $I_{1D}(q_x,L_y,L_z)$ as
\begin{equation}\label{eq:I1D}
    I_{1D}(q_x,L_y,L_z) = \int\frac{1}{q^2}\left|I_{nm}(q_y,q_z)\right|^2dq_y\;dq_z.
\end{equation}
After substituting Eq. (\ref{eq:I1D}) into Eq. (\ref{eq:gammapop}), and converting the integration over wave vector ($k_x'$) to the integration over energy ($\mathcal{E}'$), the final POP scattering rate is written as
\begin{eqnarray}\label{POPrate1}
    \Gamma_{nm}^{pop}(k_x)&=&\frac{\mid C \mid ^2 V}{4 \pi ^2 \hbar}N_0\sqrt{\frac{m^*}{2\hbar^2}}I_{1D}(q_x,L_y,L_z)\nonumber\\
    &\times&\frac{1+2\alpha\mathcal{E}_f}{\sqrt{\mathcal{E}_f(1+\alpha\mathcal{E}_f)}}\Theta(\mathcal{E}_f),
\end{eqnarray}
where $q_x=k_x\pm k_x'$ is the optical phonon wave vector along the NW axis. $\mathcal{E}_f$ is the final electron kinetic energy, which is given by
\begin{equation}
    \mathcal{E}_f=\mathcal{E}_n-\mathcal{E}_m+\mathcal{E}_i\pm\hbar\omega_0.
\end{equation}

\section{Piezoelectric Scattering}
\label{sec:PZ}

The creation of a built-in electric field by strain is called the piezoelectric effect, and this field causes piezoelectric scattering of charge carriers. Here, we show a detailed derivation of the piezoelectric scattering rate in GaN NWs. The purturbing Hamiltonian due to the piezoelectric effect is given by
\begin{equation}
    H_{pz}=\sum_q \frac{e e_{pz}^*}{\epsilon_{\infty}}\sqrt{\frac{\hbar}{2\rho V \omega_q}}\left(a_qe^{i\vec{q}\cdot\vec{r}}-a_q^{\dagger}e^{-i\vec{q}\cdot\vec{r}}\right),
\end{equation}
where $e_{pz}^*$ and $\epsilon_{\infty}$ are the effective piezoelectric constant and the high-frequency effective dielectric constant, respectively.

The matrix element for scattering from the initial electronic state $|k_x,n\rangle$ to the final state $|k_x',m\rangle$ is given by
\begin{eqnarray}\label{eq:MnmPz}
    M_{nm}(k_x,k_x')&=&\frac{ee_{pz}^*}{\epsilon_{\infty}}\sqrt{\hbar}{2\rho V\omega_q}\sqrt{N_q} \nonumber\\
    &\times&\int\left[\psi_n(y,z)e^{i(q_yy+q_zz)}\psi_m(y,z)\right]dy\;dz\nonumber\\
    &\times&\frac{1}{L_x}\int e^{i(k_x-k_x'\mp q_x)x}dx,
\end{eqnarray}
where we used the equipartition approximation for the acoustic phonon population, $N_q\simeq N_q+1\simeq\frac{k_BT}{\hbar \omega_q}$.

By assuming the linear dispersion relation for acoustic phonons, i.e. $\omega_q=\upsilon_sq$, Eq. (\ref{eq:MnmPz}) yields
\begin{eqnarray}
    \left|M_{nm}(k_x,k_x')\right|^2&=&K_{av}^2\frac{e^2k_BT}{2V\epsilon_{\infty}}\frac{1}{q^2}\\
    &\times &\left|I_{nm}(q_y,q_z)\right|^2\delta(k_x-k_x'\mp q_x),\nonumber
\end{eqnarray}
where $I_{nm}(q_y,q_z)$ is the overlap integral defined in Eq.(\ref{eq:Inm}). $K_{av}$ is called the electromechanical coupling coefficient and is defined as
\begin{equation}
    K_{av}=\sqrt{\frac{{e_{pz}^*}^2}{\rho \upsilon_s^2\epsilon_{\infty}}}.
\end{equation}

By perusing an integration procedure similar to the one done for the calculation of POP scattering rate, the piezoelectric scattering rate can be written as
\begin{eqnarray}
    \Gamma_{nm}^{pz}(k_x)&=&\frac{K_{av}^2}{4\pi^2\hbar}\frac{e^2k_BT}{\epsilon_{\infty}}\sqrt{\frac{m^*}{2\hbar^2}}I_{1D}(q_x,L_y,L_z)\nonumber\\
    &\times&\frac{1+2\alpha\mathcal{E}_f}{\sqrt{\mathcal{E}_f(1+\alpha\mathcal{E}_f)}}\Theta(\mathcal{E}_f).
\end{eqnarray}
$q_x$ is the acoustic phonon wave vector along the wire axis. The PZ scattering is an elastic process and the finite kinetic energy of electron given by $\mathcal{E}_f=\mathcal{E}_n-\mathcal{E}_m+\mathcal{E}_i$.

\end{document}